\documentclass[reprint,amsmath,amssymb,aps,prapplied,superscriptaddress]{revtex4-2}

\usepackage{hyperref}
\usepackage{graphicx}
\usepackage[alsoload=hep]{siunitx}
\usepackage{float}
\usepackage{physics}
\usepackage{makecell}
\usepackage{color}
\usepackage{outlines}
\usepackage{multirow}
\usepackage{bm}

\DeclareUnicodeCharacter{2009}{\,}

\begin{document}

\newcommand{\red}[1]{ \textcolor{red}{#1}}

\title{Optically enhanced electric field sensing using nitrogen-vacancy ensembles}

\author{M. Block}
\thanks{These authors contributed equally to this work.}
\affiliation{Department of Physics, University of California, Berkeley, CA 94720, USA}
\author{B. Kobrin}
\thanks{These authors contributed equally to this work.}
\affiliation{Department of Physics, University of California, Berkeley, CA 94720, USA}
\affiliation{Materials Science Division, Lawrence Berkeley National Laboratory, Berkeley, California 94720, USA}
\author{A. Jarmola}
\thanks{These authors contributed equally to this work.}
\affiliation{Department of Physics, University of California, Berkeley, CA 94720, USA}
\affiliation{U.S. Army Research Laboratory, Adelphi, Maryland 20783, USA}
\author{S. Hsieh}
\affiliation{Department of Physics, University of California, Berkeley, CA 94720, USA}
\affiliation{Materials Science Division, Lawrence Berkeley National Laboratory, Berkeley, California 94720, USA}
\author{C. Zu}
\affiliation{Department of Physics, University of California, Berkeley, CA 94720, USA}
\author{N. L. Figueroa}
\affiliation{Institute of Physics, Pontificia Universidad Catolica de Chile, Santiago, Chile}
\affiliation{Helmholtz Institut Mainz, Johannes Gutenberg Universitat Mainz, 55128 Mainz, Germany}
\author{V. M. Acosta}
\affiliation{Center for High Technology Materials and Department of Physics and Astronomy, University of New Mexico, Albuquerque, New Mexico 87131, USA}
\author{J. Minguzzi}
\affiliation{Institute for Quantum Electronics, ETH Zurich, 8093 Zurich, Switzerland}
\author{J. R. Maze}
\affiliation{Institute of Physics, Pontificia Universidad Catolica de Chile, Santiago, Chile}
\author{D. Budker}
\affiliation{Department of Physics, University of California, Berkeley, CA 94720, USA}
\affiliation{Helmholtz Institut Mainz, Johannes Gutenberg Universitat Mainz, 55128 Mainz, Germany}
\author{N. Y. Yao}
\affiliation{Department of Physics, University of California, Berkeley, CA 94720, USA}
\affiliation{Materials Science Division, Lawrence Berkeley National Laboratory, Berkeley, California 94720, USA}

\date{\today}

\begin{abstract}
Nitrogen-vacancy (NV) centers in diamond have shown promise as inherently localized electric-field sensors, capable of detecting individual charges with nanometer resolution.
Working with NV ensembles, we demonstrate that a detailed understanding of the  \emph{internal} electric field environment enables enhanced sensitivity in the detection of \emph{external} electric fields.
We follow this logic along two complementary paths.
First, using excitation tuned near the NV's zero-phonon line, we perform optically detected magnetic resonance (ODMR) spectroscopy at cryogenic temperatures in order to precisely measure the NV center's excited-state susceptibility to electric fields.
In doing so, we demonstrate that the characteristically observed contrast inversion arises from an interplay between spin-selective optical pumping and the NV centers' local charge distribution.
Second, motivated by this understanding, we propose and analyze a method for optically-enhanced electric-field sensing using NV ensembles; we estimate that our approach should enable order of magnitude improvements in the DC electric-field sensitivity.
\end{abstract}

\maketitle

\section{Introduction} \label{sec:intro}

The precision measurement of electric fields remains an outstanding challenge at the interface of fundamental and applied sciences \cite{hall2012high,jin2012single,huang2015imaging,dhar2014direct,ben-shach_detecting_2015}.
Leading electric field sensors are often based upon nanoelectronic systems \cite{schoelkopf_radio-frequency_1998,tosi_design_2019,johnson_singlet-triplet_2005,vamivakas_nanoscale_2011}, electromechanical resonators \cite{cleland1998nanometre,bunch2007electromechanical}, or Rydberg-atom spectroscopy \cite{fan_atom_2015,facon_sensitive_2016,kumar_atom-based_2017}.
While such techniques offer exquisite sensitivities, their versatility can be limited by intensive fabrication, calibration or operation requirements.

More recently, quantum sensors based on solid-state spin defects have emerged as localized probes \cite{awschalom_quantum_2018,dolde2011electric,taylor_high-sensitivity_2008,wolfowicz2019Heterodyne}, offering nanoscale spatial resolution and the ability to operate under a wide variety of external conditions~\cite{amsuss2011cavity,toyli2012measurement,doherty2014electronic,lesik2018magnetic,yip_measuring_2019,hsieh2018imaging,kucsko2013nanometre,schirhagl2014nitrogen}. 
The spin sub-levels of such defects are naturally coupled to magnetic fields~\cite{taylor_high-sensitivity_2008, hsieh2018imaging, kraus_magnetic_2014}, but  exhibit comparatively  weak susceptibilities to electric fields \cite{van1990electric}.
To this end, a tremendous amount of effort has focused on developing techniques to improve spin-defect-based electrometry~\cite{dolde2011electric,chen_high-sensitivity_2017,michl_robust_2019,klimov2014electrically,falk2014electrically,iwasaki2017direct, ariyaratne2018nanoscale, khivrich2019electric, anderson_electrical_2019, zhang_single_2019}.

Broadly speaking, these efforts can  be divided into two categories: (i) leveraging orbital states (as opposed to spin states), which exhibit significantly stronger coupling to electric fields, or (ii) utilizing high-density ensembles, which enhances the
sensitivity as $\sim 1/\sqrt{N}$, the standard quantum limit \cite{giovannetti2004quantum}.
Each of these approaches, however, faces its own obstacles. 
In the first case, accurate measurements of the electronic susceptibilities have proven challenging due to the deleterious effects of local charge traps observed in single defect experiments \cite{tamarat_stark_2006, acosta_dynamic_2012, PhysRevLett.107.266403, ruhl_stark_2020, white_static_2020, noh_stark_2018, muller_wide-range_2011, nagy_high-fidelity_2019, zhang_single_2019, de_las_casas_stark_2017}.
In the latter case, higher densities exacerbate inhomogeneous broadening, which can ultimately overwhelm any statistical improvement in sensitivity.

\begin{figure}
    \centering
    \includegraphics[scale=0.5]{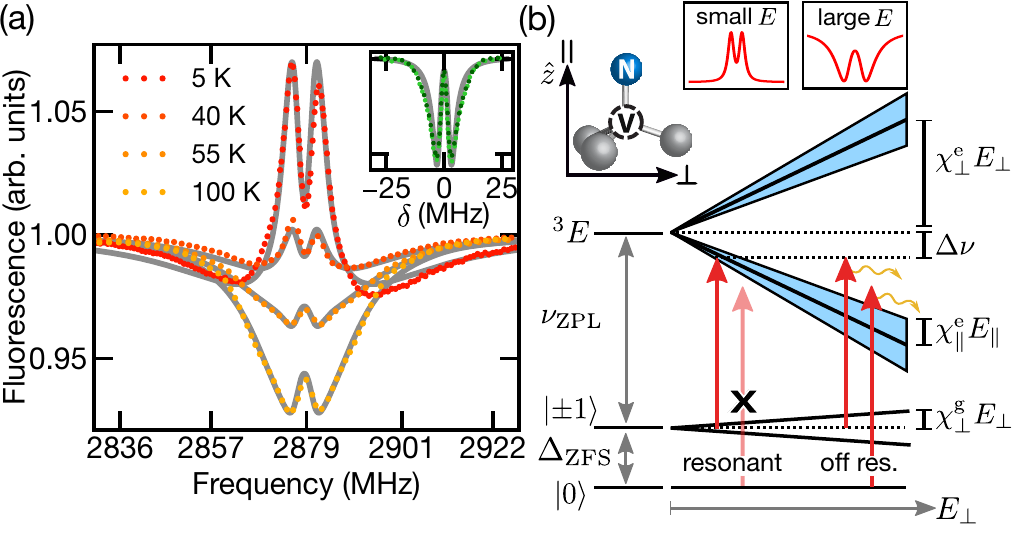}
    \caption{\small (a) Resonant ODMR at varying temperatures with drive detuning $\Delta \nu \approx 156~\si{\giga \hertz}$ below the ZPL. For $T \lesssim 45~\si{\kelvin}$, where the optical transition linewidth is smaller than $\Delta_\textrm{ZFS}$ \cite{fu_observation_2009}, we observe the emergence of sharp positive-contrast peaks \cite{akhmedzhanov_optically_2016}. Our numerical charge-based model (gray lines) quantitatively reproduces the experimental spectra. (Inset) The lineshape of the off-resonant ODMR as a function of $\delta$, the microwave detuning from $\Delta_\textrm{ZFS}$, at room temperature (dark green) and $5~\si{\kelvin}$ (light green) exhibits no temperature dependence. Resonant and off-resonant ODMR were performed at $0$ magnetic field. Error bars are smaller than the marker size. (b) NV level structure in the presence of internal electric fields. The wavelength of the ZPL transition is approximately $637.2~\si{\nano \meter}$, and resonant (off-resonant) ODMR is performed with an excitation wavelength of $636-639~\si{\nano \meter}$ ($532~\si{\nano \meter}$). The perpendicular field, $E_\perp = \sqrt{E_x^2+E_y^2}$, splits the $^3E$ manifold, while the parallel field, $E_\parallel = E_z$, shifts it (shaded blue region); the $^3E$ fine structure (not shown) is much smaller than these effects. Only perpendicular fields, which split $\ket{m_s=\pm 1}$ by $\chi^\mathrm{e}_\perp E_\perp$, strongly affect the ground state \cite{van1990electric}. Internal electric fields determine whether a given NV is: (i) \emph{resonantly} driven (favored at small $E$), resulting in positive-contrast peaks, or (ii) \emph{off-resonantly} driven (favored at large $E$), resulting in a negative contrast ODMR lineshape.}
\label{fig:1}
\end{figure}

In this paper, we propose and analyze a technique, inspired by atomic saturation spectroscopy, designed to mitigate these challenges~\cite{kehayias_microwave_2014,schawlow_spectroscopy_1982,schawlow_spectroscopy_1982}. 
In particular, we focus on dense ensembles of nitrogen vacancy (NV) color centers in diamond---a spin defect which can be optically polarized and coherently manipulated via microwave fields~\cite{doherty2013nitrogen,schirhagl2014nitrogen}.
The essence of our approach is to apply \emph{resonant} optical excitation to polarize a subgroup of an inhomogeneously broadened ensemble, and to probe the ground-state properties of this subgroup using optically detected magnetic resonance (ODMR).
In doing so, we observe an unusual spectral feature --- inverted-contrast peaks \cite{akhmedzhanov_optically_2016, akhmedzhanov_microwave-free_2017} --- which are significantly narrower than the magnetic spectra obtained via conventional, \emph{off-resonant} ODMR (Fig.~\ref{fig:1}); crucially, this feature reveals an underlying correlation between the excited- and ground-state energy levels,  which arises from the presence of internal electric fields within the diamond lattice \cite{mittiga2018imaging, bluvstein2019identifying, kolbl_determination_2019}. 

Investigating these correlations yields three main results. 
First, we develop a microscopic model for the charge-induced, electric field environment that quantitatively reproduces all features of the resonant ODMR spectra [Fig.~\ref{fig:1}(a)]. 
Second, we demonstrate the first zero-field, \emph{ensemble}-based method  to determine the NV's excited-state susceptibilities, yielding the transverse and longitudinal susceptibilities as $\chi^\mathrm{e}_\perp = 1.4 \pm 0.1 ~\si{\mega\hertz/ (\volt/\centi \meter)}$ and $\chi^\mathrm{e}_\parallel = 0.7 \pm 0.1~\si{\mega\hertz/ (\volt/\centi \meter)}$, respectively.
Third, based on our microscopic insights, we propose and analyze an electrometry protocol that combines resonant optical excitation \cite{akhmedzhanov_microwave-free_2017} with the excited-state's strong electric-field susceptibility to enable a significant improvement in expected sensitivity.
In particular, at low temperatures ($\lesssim 45 \si{\kelvin}$) we estimate a DC sensitivity of $\eta \approx1.3 \pm 0.3~\si{\milli \volt / \centi\meter / \sqrt{\hertz}}$, representing a two order of magnitude improvement compared to the best known NV methods \cite{chen_high-sensitivity_2017, !!_footnote_compare}.

\section{Inverted ODMR contrast} \label{sec:inverted-contrast}

\subsection{Overview} \label{subsec:inverted-contrast-overview}

The NV center hosts an electronic spin-triplet ground state, where, in the absence of perturbations, the $\ket{m_s=\pm 1}$ sublevels are degenerate and sit $\Delta_{\textrm{ZFS}} = 2.87$~GHz above the $\ket{m_s=0}$ state [Fig.~\ref{fig:1}(b)].
In high-density NV ensembles, this degeneracy is lifted most strongly by the local charge environment which directly couples the $\ket {m_s = \pm 1}$ sublevels; this leads to typical ground-state ODMR spectra which exhibit a pair of heavy-tailed resonances centered around $\Delta_{\textrm{ZFS}}$ [inset, Fig.~\ref{fig:1}(a)] \cite{mittiga2018imaging,1_footnote_strain}.

Such ODMR spectra are usually obtained using continuous-wave \emph{off-resonant} optical excitation, in which the NV center is initialized and read out with laser frequency detuned far above the zero-phonon line (ZPL), $\nu_{\textrm{ZPL}}$ [Fig.~\ref{fig:1}(b)]. 
During such off-resonant excitation, the $\ket{m_s= \pm1}$ states fluoresce less brightly than the $\ket{m_s=0}$ state.
Moreover, in the absence of microwave excitation, the NV population accumulates in the $\ket{m_s = 0}$ state, owing to the spin-selective branching ratio of the singlet-decay channel.
Applying resonant microwave excitation thus drives the population from the (brighter) $\ket{m_s = 0}$ state to the (dimmer) $\ket{m_s = \pm 1}$ states.
This leads to the typically observed negative-contrast ODMR feature [inset, Fig. \ref{fig:1}(a)].
We note that for off-resonant driving, identical spectra are observed at both room temperature and cryogenic conditions [inset, Fig. \ref{fig:1}(a)].

In contrast, continuous-wave ODMR spectra taken with an optical drive near resonance with the ZPL exhibit a marked temperature dependence characterized by two principal features [Fig. \ref{fig:1}(a)].
Most prominently, for temperatures $\lesssim45~\si{\kelvin}$, the resonances \emph{invert}, becoming a pair of narrow, positive-contrast peaks \cite{akhmedzhanov_optically_2016}.
The entire spectrum, however, does not invert: Rather, these sharp peaks sit inside a broad envelope of negative contrast which is relatively temperature independent.

\begin{figure}
    \includegraphics[scale=0.5]{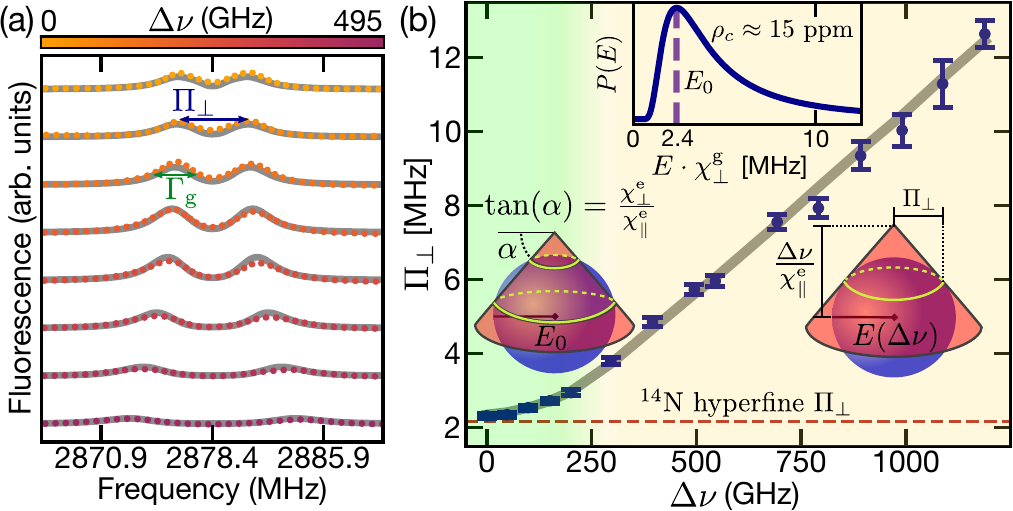}
    \caption{\small (a) Resonant ODMR spectra as a function of the detuning below ZPL, $\Delta \nu$, taken under a $20~\si{\gauss}$ magnetic field perpendicular to the NV axis at $8~\si{\kelvin}$. The positive-contrast peaks in the spectra are characterized by a splitting, $\Pi_\perp$, and a linewidth $\Gamma_\mathrm{g}$. Gray lines correspond to our numerical model. Error bars are smaller than the marker size. (b) $\Pi_\perp$ as a function of $\Delta \nu$. In the small detuning region (light green), the highest-probability electric-field sphere (blue) intersecting the resonant cone (red) is of radius $E_0$ \cite{5_footnote_intersections}; in the large detuning region (light yellow), the radius of the  highest-probability sphere that interesects the cone depends linearly on $\Delta \nu$. The red dashed line indicates the limit to $\Pi_\perp$ imposed by the hyperfine interaction. By fitting our numerical model to this data (gray line), we extract the excited-state electric-field susceptibilities. (upper inset) The probability distribution $P(E)$ exhibits a peak at $E_0$, which is determined by the charge density.}
    \label{fig:2}
\end{figure}

To understand the coexistence of these features, one must consider the interplay between resonant optical pumping and the local charge environment \cite{akhmedzhanov_optically_2016, mittiga2018imaging}.
Under resonant optical excitation, only one of the ground-state sublevels is driven to the excited state [Fig.~\ref{fig:1}(b)], while the other sublevels are optically dark and hence accumulate population. 
Microwave excitation drives population back into the resonant sublevel, leading to an increase in flourescence --- i.e.~a positive-contrast ODMR feature [Fig.~\ref{fig:2}(a)].
This resonant pumping mechanism is highly dependent on temperature: it can only occur if the thermally-broadened optical transition linewidth is smaller than $\Delta_{\textrm{ZFS}}$, a situation that arises for $T \lesssim 45~\si{\kelvin}$ \cite{fu_observation_2009}.

The above picture is complicated by the presence of internal electric fields, which perturb the NV's excited-state energy levels, leading to a distribution of optical resonance conditions within the NV ensemble.
In particular, perpendicular electric fields (relative to the NV axis) split both the excited-state manifold and the ground-state $\ket {m_s = \pm 1}$ sublevels [Fig.~\ref{fig:1}(b)] \cite{7_foonote_config}.
Crucially, this \emph{correlates} the optical resonance condition and the ground-state splitting.
Indeed, for relatively small optical detunings (Fig.~\ref{fig:1}), the resonance condition is generally satisfied by NVs subject to weak local electric fields; hence, the positive-contrast feature is relatively sharp and narrowly split.
Meanwhile, the off-resonant pumping mechanism is more likely for NVs subject to large electric fields, resulting in a broad, negative-contrast background [Fig.~\ref{fig:1}(a)].
It is the superposition of these two features that gives rise to the unusual lineshapes observed in Fig.~\ref{fig:1}(a).

\subsection{Microscopic model} \label{subsec:inverted-contrast-model}

Let us now turn our heuristic understanding into a quantitative microscopic model which takes into account: ($i$) the electric field distribution, ($ii$) the excited state resonance condition, and ($iii$) the ODMR lineshape for individual NV centers under resonant and off-resonant conditions.

%
%
To begin, we consider the internal electric field distribution $P(\vec E)$ arising from randomly placed elementary charges at an overall density $\rho$.
Physically, we expect these charges to consist primarily of the NV centers themselves (which are electron acceptors) and their corresponding donors --- hence, $\rho \approx 2 \rho_{\rm{NV}}$, where $\rho_{\rm{NV}}$ is the NV defect density \cite{mittiga2018imaging}.
As the angular distribution of $\vec E$ is fully symmetric, it suffices to consider the distribution for the electric field strength, $P(E)$.
In Appendix \ref{app:model}, we demonstrate via Monte Carlo simulations that this distribution may be approximated by the analytic expression, 
\begin{align} \label{eq:P(E)}
    P_E(\tilde{E}) \; d\tilde{E} &= \frac{4 \pi}{\tilde{E}^{5/2}} \exp{-\frac{4 \pi}{3 \tilde{E}^{3/2}}} d\tilde{E}.
\end{align}
Here, $\tilde E = E/E_\textrm{ref}$ is a dimensionless electric field, where $E_\mathrm{ref} = (2 \rho)^{2/3}/(4 \pi \epsilon_0 \epsilon_r)$ is approximately the electric field strength of the nearest charge, $\epsilon_0$ is the vacuum permittivity, and $\epsilon_r$ is the relative permittivity of diamond, which we take to be 5.7 \cite{bhagavantam_dielectric_1948}.

Second, we consider the optical resonance condition given by the energy levels of the $^3$E excited state.
In particular, we assume that electric fields, which couple directly to the orbital degree of freedom, dominate over hyperfine effects, including spin-orbit coupling and spin-spin interactions (see Appendix~\ref{app:electrometry}) \cite{maze_properties_2011}. 
It is thus sufficient to model the  excited state as two branches (upper and lower) of states, whose energies relative to $\nu_\textrm{ZPL}$  are given by \cite{maze_properties_2011,doherty_negatively_2011}
\begin{align}
\label{eq:res-cond}
    \Delta \nu_{\mathrm{U}, \mathrm{L}}(\vec E) =  \chi^\mathrm{e}_\parallel E_\parallel \mp \chi^\mathrm{e}_\perp E_\perp \,
\end{align}
Note that in our notation positive detuning is \emph{below} the ZPL.
The resonance condition for a given NV center optically excited with a laser detuning $\Delta \nu$ is then given by the function,
\begin{align}
 D_R(\vec E,\Delta \nu) &= \left[\Theta(\gamma_\mathrm{e}/2 - \abs{\delta_\mathrm{U}}/2) + \Theta(\gamma_\mathrm{e}/2 - \abs{\delta_\mathrm{L}}/2)\right] \\
    \delta_\textrm{U,L} &= \Delta \nu - \Delta \nu_\textrm{U,L}(\vec E)\ ,
\end{align}
where $\gamma_\mathrm{e}$ is the single-NV linewidth of the optical transition 
and $\Theta$ is the Heaviside step function.
In particular, $D_R(\vec E,\Delta \nu)$ is 1 on resonance and 0 otherwise.

Finally, we model the ground-state ODMR of each \emph{single} NV using a primitive lineshape.
This lineshape, denoted $\Lambda(\omega;E_\perp)$, is parameterized by the perpendicular electric field $E_{\perp}$, which determines the splitting between the $\ket{m_s = \pm 1}$ sublevels.
It also incorporates two forms of broadening: ($i$) magnetic broadening arising from nearby spins (e.g.~nitrogen defects and $^{13}\rm{C}$ nuclear spins), and ($ii$) non-magnetic broadening, which includes microwave power broadening and strain.
The explicit form of $\Lambda(\omega;E_\perp)$ is provided in Appendix \ref{app:model}. 

Putting all this together, we now determine the \emph{ensemble} resonant ODMR.
This consists of two separate contributions.
The first is due to resonantly driven NVs and is given by integrating over primitive lineshapes whose associated electric field matches resonance condition:
\begin{align}
    S_\mathrm{R}(\omega;\Delta \nu) = \int dE \; P(E) \int \sin(\theta) d\theta & \Lambda(\omega; E_\perp)
    D_R(\vec E,\Delta \nu)\ ,
\end{align}
where $\vec E = (E_\perp,E_\parallel) = (E \sin \theta,E\cos \theta)$.
An analogous expression (see Appendix \ref{app:model}) describes the contribution due to off-resonantly driven NV centers.
Adding these two cases together with a relative contrast factor, $\epsilon_C$, yields the full spectrum:
\begin{align}
    S_\textrm{tot}(\omega;\Delta \nu) = \epsilon_C S_\mathrm{R}(\omega;\Delta \nu) - S_\mathrm{OR}(\omega;\Delta \nu).
\end{align}
The sign of $\epsilon_C$ determines whether the resonantly driven NV centers exhibit positive or negative contrast, while its magnitude depends on the details of the resonant optical pumping mechanism.

Using the above model, we perform numerical simulations of both resonant and off-resonant ODMR spectra for a range of temperatures.
While our simulations depend on several input parameters, the most physically relevant of these are constrained by independent analysis. 
In particular, we determine the charge density $\rho \approx 15 \pm 2$~ppm by fitting the \emph{off-resonant} ODMR spectra to our charge-based model [inset, Fig.~\ref{fig:1}(a)] \cite{mittiga2018imaging}; this suggests an NV density of $\rho_{\textrm{NV}} \approx \rho/2 \approx 8$ ppm, which is consistent with prior density estimates for this sample \cite{acosta_diamonds_2009}.
In addition, as we discuss at length in the following section, we determine the excited state electric-field susceptibilities from independent measurements of resonant ODMR as function of optical detuning.
The remaining temperature-dependent fit parameters, related to linewidth broadening and relative ODMR contrast, are provided in Appendix~\ref{app:temp}.

The resulting lineshapes shown in  Fig.~\ref{fig:1}(a) are in excellent agreement with the experimental data across all temperatures.
Notably, even at high temperatures where the striking positive-contrast peak is absent, the resonant experimental spectra remain qualitatively distinct from the off-resonant spectra, yet are correctly captured by our lineshape simulations. 

\section{Excited State Electric-field Susceptibilities} \label{sec:susceps}

Interestingly, the correlation between the positive-contrast peaks and the optical resonance condition inspires a means of determining the \emph{excited-state} electric-field susceptibilities from \emph{ground-state} ODMR spectroscopy.
In particular, as shown in Fig.~\ref{fig:2}(a), we perform ODMR measurements of the inverted-contrast feature as a function of the optical detuning.
By tracking how the splitting, $\Pi_\perp$, of the positive-contrast feature changes as a function of $\Delta \nu$, we fully determine the excited-state susceptibilities, $\chi^\mathrm{e}_\parallel$ and $\chi^\mathrm{e}_\perp$ [Fig.~\ref{fig:2}(b)].
At its core, this ability to independently extract the susceptibilities stems from the fact that Fig.~\ref{fig:2}(b) exhibits two distinct regimes: at small detunings, $\Pi_\perp$ exhibits a suppressed dependence on $\Delta \nu$, while at large detunings, $\Pi_\perp$ exhibits a linear dependence.

Let us now explain the origin of these two regimes.
The splitting, $\Pi_\perp$, of the positive-contrast ODMR feature is controlled by: (i) the optical resonance condition and (ii) the distribution of electric fields.
We focus on resonance with the lower branch, which is dominant resonant pumping mechanism for optical detunings below the ZPL (see Appendix~\ref{app:above-zpl}).
This resonance condition (equation~\eqref{eq:res-cond}) can be rearranged to obtain
\begin{align} \label{eq:cone}
    E_z - \frac{\Delta \nu} {\chi^\mathrm{e}_\parallel} = - \frac{\chi^\mathrm{e}_\perp}{\chi^\mathrm{e}_\parallel} \sqrt{E_x^2 + E_y^2},
\end{align}
which defines a ``resonant cone'' in electric field space with apex at $E_z = \Delta \nu / \chi^\mathrm{e}_\parallel$ [Fig.~\ref{fig:2}(b)].
On the other hand, the electric-field distribution is spherically symmetric and peaked at a characteristic electric field, $E_0 \cdot \chi^\mathrm{g}_\perp \approx 2.4 ~\si{\mega \hertz}$, set by $\rho$ [inset, Fig.~\ref{fig:2}(b)].

For a given detuning, this provides a geometric interpretation for determining the electric field configurations most likely to match the resonance condition; in particular, these configurations are set by the highest-probability sphere that intersects the resonant cone [yellow circles in Fig.~\ref{fig:2}(b)].
At small detunings, this sphere is always at radius $E_0$, implying that $\Pi_\perp \sim E_0$  can only weakly depend on the detuning \cite{5_footnote_intersections}.
At large detunings, the sphere of radius $E_0$ no longer intersects the cone, and instead, the highest-probability intersecting sphere is the inscribed sphere [Fig.~\ref{fig:2}(b)].
The size of the inscribed sphere grows linearly with the detuning, and thus so does $\Pi_\perp$.

As a result of these two regimes, $\Pi_\perp(\Delta \nu)$ in Fig.~\ref{fig:2}(b) has both a slope, $m_\Pi$, and an elbow, at $\Delta \nu = \Delta \nu ^*$.
From this information alone, we can analytically estimate $\chi^\mathrm{e}_\parallel$ and $\chi^\mathrm{e}_\perp$.
In particular, setting $\alpha = \tan^{-1}(\chi^{\rm{e}}_\perp / \chi^{\rm{e}}_\parallel)$ be the exterior angle of the resonant cone, we obtain
\begin{align}
    \sin(\alpha) &= m_\Pi \cdot \Delta \nu^* / (E_0 \chi^\mathrm{g}_\perp) \\
    \frac{\chi^\mathrm{e}_\parallel}{\chi^\mathrm{g}_\perp} &=  \cos(\alpha) \Delta \nu^* / (E_0 \chi^\mathrm{g}_\perp).
\end{align}
We estimate $\Delta \nu ^ * = 200~\si{\giga \hertz}$ and $m_\Pi = 10^{-5}$ from Fig.~\ref{fig:2}(b) and $E_0 \chi_\perp^\textrm{g}=2.4~\si{\mega \hertz}$ from the off-resonant spectra [Fig.~\ref{fig:1}(b), inset].
This yields $\chi^\mathrm{e}_\perp \approx 1.2~\si{\mega\hertz/ (\volt/\centi \meter)}$ and $\chi^\mathrm{e}_\parallel \approx 0.8~\si{\mega\hertz/ (\volt/\centi \meter)}$. 

To refine these estimates and corroborate our geometric interpretation, we simulate the full resonant ODMR lineshape as a function of optical detuning and fit the simulated $\Pi_\perp$ to the experimental data.
Unlike the analytic estimates above, this model takes into account all resonant electric-field configurations.
To perform the fits, we fix all parameters of our microscopic model except for the susceptibilities (see Appendix~\ref{app:susceps} for details). 
The best-fit susceptibilties are given by $\chi^\mathrm{e}_\perp = 1.4 \pm 0.1 ~\si{\mega\hertz/ (\volt/\centi \meter)}$ and $\chi^\mathrm{e}_\parallel = 0.7 \pm 0.1~\si{\mega\hertz/ (\volt/\centi \meter)}$ [gray line in Fig.~\ref{fig:2}(b)], which agree within error bars with the analytic estimates.

These results represent a refinement over previous measurements of the excited state susceptibilities via single NV Starks shifts, which are strongly distorted by photo-ionized charge traps \cite{tamarat_stark_2006, acosta_dynamic_2012, PhysRevLett.107.266403}.
In contrast, ensemble measurements appear to be insensitive to the effects of charge traps  \cite{van1990electric,chen_high-sensitivity_2017,michl_robust_2019}; indeed, assuming their positions are random, charge traps would contribute to the effective charge density but would not systematically bias the ensemble Stark shift \cite{6_footnote_chargedyn}.

\begin{figure}
    \includegraphics[scale=0.5]{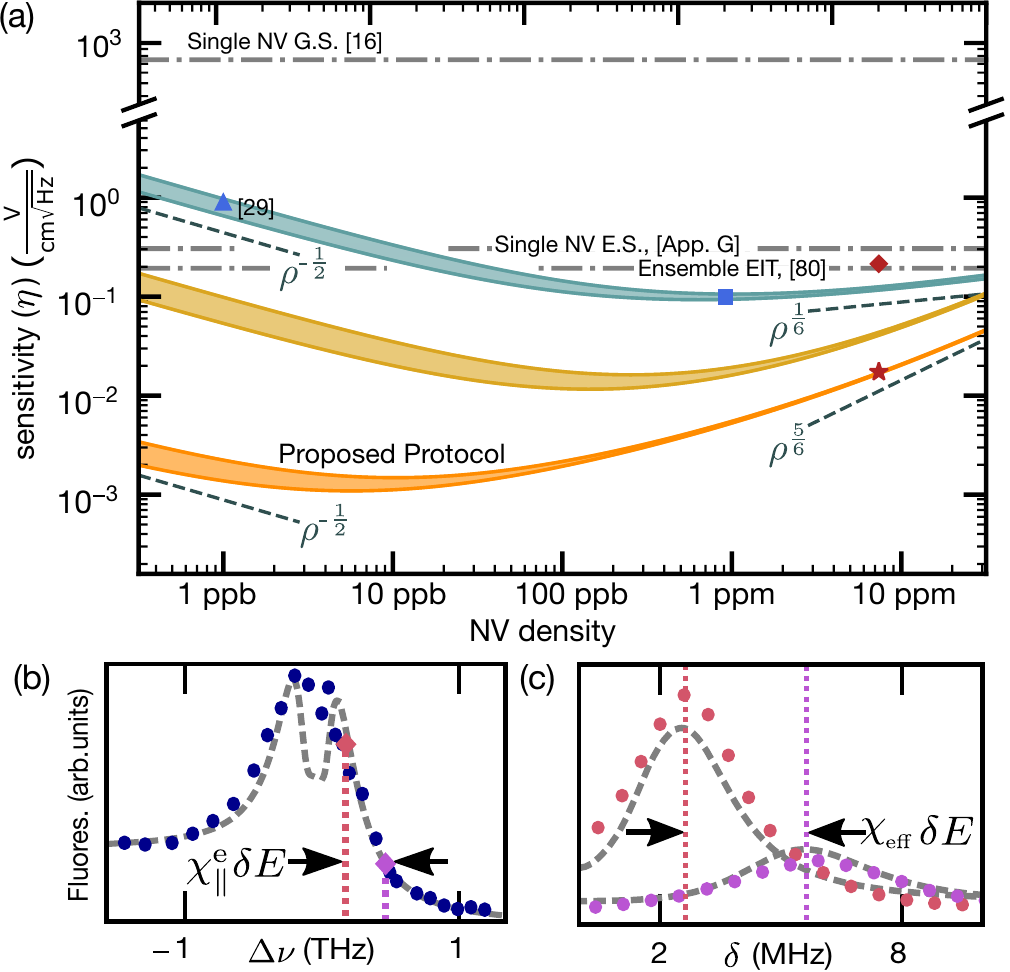}
    \caption{\small (a) Comparison of sensitivities for various NV-based DC electric field sensing methods. Teal region: estimated sensitivity using established NV ensemble electrometry techniques at an illumination volume of $0.1~\si{\milli \meter}^3$. The light-blue triangle marks the sensitivity achieved in \cite{chen_high-sensitivity_2017}; the light-blue square marks the optimal sensitivity for this method based on our scaling arguments. Additional demonstrations of NV electrometry include \cite{michl_robust_2019, sharma_imaging_2018, alghannam_measuring_2018, razeghi_electric_2017, li_nanoscale_2020}. Orange (yellow) region: estimated sensitivity for our optically-enhanced protocol assuming an excited-state broadening of $10~\si{\giga \hertz}$ ($100~\si{\giga \hertz}$), an illumination volume of  $0.1~\si{\milli \meter}^3$ $(0.015~\si{\milli \meter}^3)$, and a temperature of $\lesssim 45~\si{\kelvin}$ ($\lesssim 100~\si{\kelvin}$). Red star (diamond): estimated sensitivity for our sample  at low temperature (for the microwave-free variant of our protocol at $300~\si{\kelvin}$). For our sample, we take the paramagnetic broadening to be the experimentally measured value $\sim 1.7~ \si{\mega \hertz}$ (see Appendix~\ref{app:electrometry}). For all other other sensitivity estimates, we use a density-dependent model for the paramagnetic ODMR linewidth, assuming a natural abundance of $^{13}\mathrm{C}$ \cite{edmonds_generation_2020, barry_sensitivity_2020}. Dashed lines indicate asymptotic scaling of sensitivities. (b) Measured fluorescence as a function of $\Delta \nu$ (blue dots) and simulated fluorescence (dashed lines, see Appendix~\ref{app:electrometry} for details). An external field of strength $\delta E$ results in a change in overall fluorescence. (c) Measured peak shift of the resonant ODMR spectra (pink, purple dots) and simulated peak shift (dashed lines) for an external field of strength $\delta E$.}
    \label{fig:3}
\end{figure}

Beyond allowing us to extract the susceptibilities, our numerical model fully reproduces the detuning-dependent experimental data [Fig.~\ref{fig:2}(a)].
In particular, the model quantitatively recovers two characteristic features of these spectra: a decrease in the overall fluorescence and an increase in the linewidth $\Gamma_\mathrm{g}$, for increasing $\Delta \nu$.
Physically, fluorescence declines with $\Delta \nu$ because the larger electric fields required for resonance are less likely.
The dependence of $\Gamma_\mathrm{g}$ on $\Delta \nu$ is more subtle and is discussed in Appendix~\ref{app:linewidth}.

\section{Optically Enhanced Electrometry} \label{sec:electrometry}

Our understanding of the interplay between internal electric fields and resonant excitation suggests a protocol for DC electric field sensing using NV ensembles.
The protocol is premised on the fact that an external electric field parallel to the NV axis induces an overall shift of the excited-state levels.
In effect, this is equivalent to changing the optical detuning, which we have already observed has two primary consequences: (i) it alters the splitting of the inverted-contrast peaks [Fig.~\ref{fig:3}(c)], and (ii) it changes the density of resonant configurations and therefore the overall fluorescence [Fig.~\ref{fig:3}(b)]
\footnote{Interestingly, the first of these effects has also been explored in the context of a complementary MW-free magnetometry protocol \cite{akhmedzhanov_microwave-free_2017}}.

To leverage these effects for electrometry, we propose the following protocol.
First, apply a bias electric field parallel to one of the NV orientations to spectrally isolate its excited state \cite{3_footnote_bias}.
Second, perform resonant ODMR with fixed laser detuning below the peak of the ZPL such that positive-contrast peaks are clearly observed.
Because overall fluorescence decreases with detuning, the optimal choice is the \emph{smallest} detuning such that the positive contrast peaks disperse linearly with an applied field [see Appendix~\ref{app:electrometry}, Fig~\ref{fig:sensing-ests}(a)].
Finally, monitor the fluorescence at a fixed microwave drive frequency that maximizes the slope of the inner edge of one of the resonant ODMR peaks. 
Unless otherwise stated, we assume operating temperatures of $\lesssim 45~\si{\kelvin}$ throughout our discussion, as required for the occurrence of positive-contrast peaks. 

Unlike typical NV electric field sensing methods, our protocol is sensitive to fields \emph{parallel} to the NV axis and insensitive to perpendicular fields.
Intuitively, the insensitivity to perpendicular fields owes to the random orientation of internal electric fields.
To illustrate this, consider the level shift induced by a small perpendicular field $\delta E_\perp$ oriented in the $\hat{x}$ direction and assume internal perpendicular fields are randomly oriented in the $xy$ plane with strength $E_0$.
The ensemble-average level shift, $\delta \nu$, of the lower branch is then given by
\begin{align}
    \delta \nu/\chi^\mathrm{e}_\perp &= \frac{1}{2\pi}\int d\theta \sqrt{E_0^2 + \delta E_\perp^2 + 2\delta E_\perp E_0 \cos{\theta}} - E_0 \\
    &\sim O(\delta E_\perp^2 / E_0),
\end{align}
which vanishes at leading order in $\delta E_\perp$.

We now evaluate the sensitivity of our protocol to parallel fields.
We first estimate the sensitivity owing to the peak shift alone \cite{taylor_high-sensitivity_2008}:
\begin{equation}
\eta_{\Pi} = P_\Pi\frac{\Gamma_{\textrm{g}}}{\chi_\textrm{eff} C_0 C_\mathrm{r}}\cdot \frac 1 {\sqrt R},
\end{equation}
where $\Gamma_{\textrm{g}}$ is the linewidth of the positive-contrast peak, $P_\Pi \approx 0.77$ is a numerical factor associated with the lorentzian lineshape, $C_0 \approx 0.21$ is the inherent ODMR contrast, $R$ is the total photon count rate, and $C_\mathrm{r} \approx 0.55$ is the ratio of resonant fluorescence to total fluorescence.
Note that $\chi_\textrm{eff} \approx 0.41 \chi^\textrm{g}_\perp$ is an effective susceptibility (right inset, Fig.~\ref{fig:3}) related to the slope of $\Pi_\perp$ with respect to $\Delta \nu$ [Fig.~\ref{fig:2}(b)].
Similarly, we estimate the sensitivity due to overall fluorescence variation:
\begin{equation}
\eta_{\textrm{F}} = P_{\textrm{F}} \frac{\Gamma_{\textrm{e}}}{\chi^{\textrm{e}}_{\parallel} C_\mathrm{r}}\cdot \frac 1 {\sqrt R},
\end{equation}
where $\Gamma_{\textrm{e}}$ is the linewidth of the optical transition and $P_{\textrm{F}} \approx 0.39$ is a numerical lineshape factor determined from experimental data (see Appendix~\ref{app:electrometry}).
The change in fluorescence due to both these mechanisms may be combined, leading to an overall sensitivity: $1/\eta = 1/\eta_{\Pi} + 1/\eta_{\textrm{F}}$ (see Appendix~\ref{app:electrometry}).

For our current sample, one finds a sensitivity, $\eta = 18 \pm 4~\si{\milli \volt/\centi \meter / \sqrt \hertz}$, assuming an illumination volume of $0.1~\si{\milli \meter}^3$ \cite{chen_high-sensitivity_2017}.
This represents a $5 \times$ improvement over established NV electrometry techniques (Fig.~\ref{fig:3}).
The enhancement in sensitivity derives primarily from three factors: (i) a larger photon count rate due to resonant scattering, (ii) an improvement in contrast, and (iii) the ability to constructively combine the signal from peak-shifting and fluorescence variation. 

The sensitivity of our protocol can be further improved by optimizing the NV density.
As in our numerical model, let us assume that the total charge density is twice the NV density.
At low densities, $\eta_{\Pi}$ and $\eta_{\textrm{F}}$ are limited by the intrinsic broadening of resonant ODMR and the optical transition, respectively.
By increasing density, both sensitivities improve according to the standard quantum limit, $\eta \propto 1/\sqrt{\rho_\textrm{NV}}$ --- the usual motivation for performing ensemble sensing (Fig.~\ref{fig:3}).
However, at sufficiently high densities, the broadening due to internal electric fields becomes larger than the intrinsic broadening and the sensitivity degrades (Fig.~\ref{fig:3})~\cite{!_footnote_widthlim}; intuitively, this occurs because the NV ensemble is primarily sensing electric fields within the diamond lattice rather than the external signal.
In particular, we show in Appendix~\ref{app:electrometry} that the sensitivity degrades upon increasing density as $\eta \sim \rho_\textrm{NV}^{5/6}$ (Fig.~\ref{fig:3}).
Conversely, the sensitivity improves rapidly upon decreasing density until one reaches the crossover density between the intrinsically-broadened and charge-broadened regimes.

Interestingly, this crossover density is naturally different for $\eta_\Pi$ and $\eta_\mathrm{F}$.
In particular, the non-charge-induced broadening of the \emph{ground-state} ODMR linewidth is often limited to $\sim 200~\si{\kilo \hertz}$ by the $^{13}$C nuclear spin bath (although isotopically purified samples can exhibit narrower linewidths, changing the crossover density; this is discussed in more detail in Appendix~\ref{app:electrometry})~\cite{linh-pham-thesis, balasubramanian_ultralong_2009, edmonds_generation_2020}.
This implies that $\eta_\Pi$ is optimal at NV densities of $\sim 30$ ppb.
On the other hand, these same magnetic fields only weakly affect the excited state.
Rather, the non-charge-induced broadening of the excited-state, whose origin is less well understood, has been empirically observed to be $\sim 10~\si{\giga \hertz}$ \cite{batalov_low_2009, PhysRevLett.110.213605, 9_footnote_eswidth}.
This yields an optimal NV density for $\eta_\mathrm{F}$ of $\sim 7$ ppb.

Putting everything together, we obtain an optimal total sensitivity of $\eta = 1.3 \pm 0.3~\si{\milli \volt / \centi\meter / \sqrt{\hertz}}$ at an NV density $\sim 10$ ppb (Fig.~\ref{fig:3}, Table~\ref{tab:electrometry}).
This represents a two order of magnitude enhancement compared state-of-the-art NV methods (though these do not require cryogenic temperatures \cite{chen_high-sensitivity_2017,!!_footnote_compare}).

A few remarks are in order.
First, while our sensitivity estimates assume an optically-thin sample, comparable sensitivities may be achieved at larger optical depths by monitoring, for example, transmission amplitude instead of fluorescence \cite{PhysRevLett.110.213605}.
Second, monitoring resonant fluorescence variation alone via resonant excitation --- without performing ODMR --- already provides a significant electric field sensitivity, yielding a microwave-free version of our protocol.
Since this microwave-free protocol does not require one to track the positive-contrast ODMR feature, it can also be applied at room temperature [Fig.~\ref{fig:1}(a)].
Assuming a thermally broadened linewidth of $\sim 2~\si{\tera \hertz}$ at $300~\si{\kelvin}$ yields a sensitivity of $\approx 300~\si{\milli \volt / \centi \meter / \sqrt \hertz}$ (Fig.~\ref{fig:3}); this is comparable to the best reported NV sensitivities at room temperature.
Relatedly, our protocol may be extended to radiofrequency electrometry through Fourier analysis of the time-dependent fluorescence \cite{shin_room-temperature_2012}.

\section{Conclusion} \label{sec:conclusion}

Our work opens the door to a number of intriguing future directions.
First, in combination with recent work on diamond-surface-termination \cite{oberg2019solution,cui2013increased}, our protocol's enhanced sensitivities may help to mitigate the deleterious effects of surface screening, which currently limit the NV's ability to detect external electric fields \cite{broadway_spatial_2018, stacey_evidence_2019, mertens_patterned_2016}.
Second, our spectroscopy tools are generically applicable to characterizing the charge environment in defect systems.
In particular, non-linear Stark shifts, consistent with the presence of local charges, have been observed in a multitude of defects, including: boron-vacancy in \emph{h}-BN, chromium in diamond, and both silicon-vacancy and divacancies in 4H-SiC \cite{noh_stark_2018, gottscholl_initialization_2020, sajid_edge_2020, muller_wide-range_2011, ruhl_stark_2020, white_static_2020, nagy_high-fidelity_2019, de_las_casas_stark_2017}.
These non-linear Stark effects hinder the accurate experimental determination of  susceptibilities, making it challenging to assess the potential of such defect systems  for quantum metrology.
Finally, our sensitivity scaling analysis suggests that for any defect ensemble  exhibiting charge-dominated, inhomogeneous broadening, one can dramatically optimize electric-field sensitivities by carefully tuning the defect density. 
Such enhanced sensitivities could enable the observation of new quantum transport phenomena \cite{belthangady2013dressed,oberg2019spin} as well as mesoscopic quantum thermodynamics studies \cite{mohammady2018low}. 

\emph{Acknowledgements.---}We gratefully acknowledge the insights of and discussions with A.~Norambuena, T.~Mittiga, P.~Maletinsky, A.~Jayich, and H.~Zheng.
We thank D.~Suter and W.~Wu for a careful reading of the manuscript.
We are especially grateful to K. M. Fu for sharing her raw data on the optical transition linewidth vs temperature.
This work was supported as part of the Center for Novel Pathways to Quantum Coherence in Materials, an Energy Frontier Research Center funded by the U.S. Department of Energy, Office of Science, Basic Energy Sciences under Award No.~DE-AC02-05CH11231.
M.B.~ acknowledges support through the Department of Defense (DoD) through the National Defense Science \& Engineering Graduate (NDSEG) Fellowship Program.
A.J.~acknowledges support from the Army Research Laboratory under Cooperative Agreement no.~W911NF-16-2-0008.
S.H.~acknowledges support from the National Science Foundation Graduate Research Fellowship under grant no.~DGE-1752814.
J.R.M.~acknowledges support from ANID-Fondecyt grant 1180673, AFOSR FA9550-18-1-0513 and ANID-PIA ACT102023.
This work of D.B.~was supported in part by the EU FET OPEN Flagship Project ASTERIQS.

\appendix

\section{Experimental Setup} \label{app:setup}

\begin{figure}[t]
   \centering
   \includegraphics[scale=0.45]{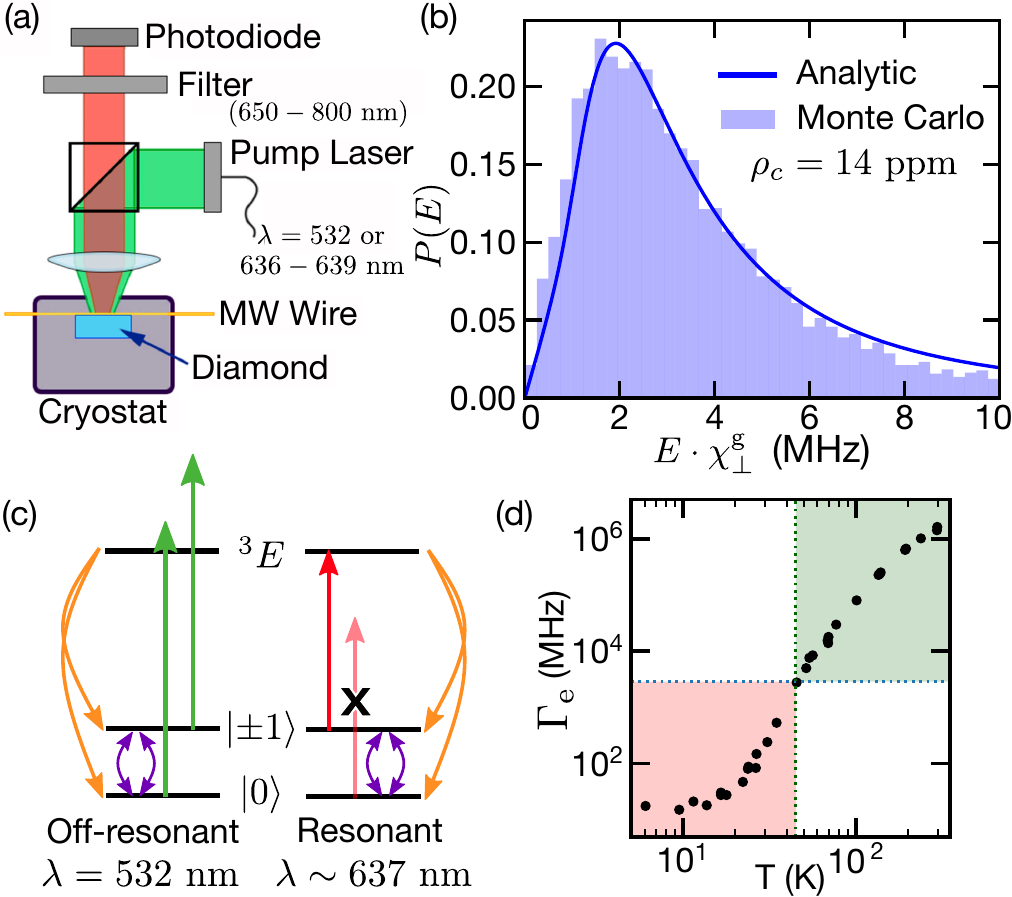}
   \caption{(a) Experimental setup. Both green and red lasers can be used for NV excitation. (b) Comparison of analytic (solid line) and Monte Carlo (histogram) models of $P(E)$. They are in close agreement, particularly near the peak of $P(E)$. (c) Off-resonant (left) and resonant (right) excitation schemes. Under resonant excitation, an effective dark state ($\ket{0}$ above) results in \emph{increased} fluorescence on microwave resonance. (d) Optical transition linewidth ($\Gamma_\mathrm{e}$) as a function of temperature. Data reproduced from \cite{fu_observation_2009}. In the green-shaded region, $\Gamma_\mathrm{e} > \Delta_\mathrm{ZFS}$ and no inverted contrast is observed; in the red-shaded region $\Gamma_\mathrm{e} < \Delta_\mathrm{ZFS}$ and resonant excitation yields inverted contrast ODMR.}
   \label{fig:background}
\end{figure}

The experimental apparatus is illustrated in Fig.~\ref{fig:background}(a). A resonant (636-639 \si{\nano\meter}, 0.2 \si{\milli \watt}) or off-resonant (532 \si{\nano\meter}, $\sim 1~\si{\milli \watt}$) laser light is focused with a 0.5 numerical-aperture, 8 \si{\milli\meter} focal length aspheric lens onto the surface of a (111)-cut diamond housed in a continuous-flow cryostat (Janis ST-500).
Fluorescence was collected using the same lens, spectrally filtered (within $650-800$ \si{\nano\meter}), and detected with a Si photodiode.
Microwaves were delivered by a $75$ $\mu$m diameter copper wire running across the surface of the diamond.
The temperature was measured with a diode located at the base of the cryostat's sample holder.

The diamond used in this work, labeled S2 in \cite{acosta_diamonds_2009}, was grown under high-pressure-high-temperature conditions (HPHT) and initially contained $\sim 100$ ppm of substitutional nitrogen.
It was then irradiated with $3$ \si{\mega\electronvolt} electrons at a dose of $10^{19}$ \si{\centi\meter^{-2}} in order to produce a uniform distribution of vacancies, and subsequently annealed at $1050$ \si{\celsius} for two hours in order to facilitate the formation of NV centers by mobilizing the vacancies.
After this treatment, the sample contains $\sim 16$ ppm of NV$^-$ and $\sim 50$ ppm of unconverted substitutional nitrogen or NV$^0$ based on ZPL intensity measurements \cite{acosta_diamonds_2009}.
We note that this estimate of the NV density is $\sim 2\times$ larger than that of the charge-based model (see subsection ~\ref{subsec:inverted-contrast-model}).

\section{Detuning Above ZPL} \label{app:above-zpl}
\begin{figure}
   \centering
   \includegraphics[scale=0.45]{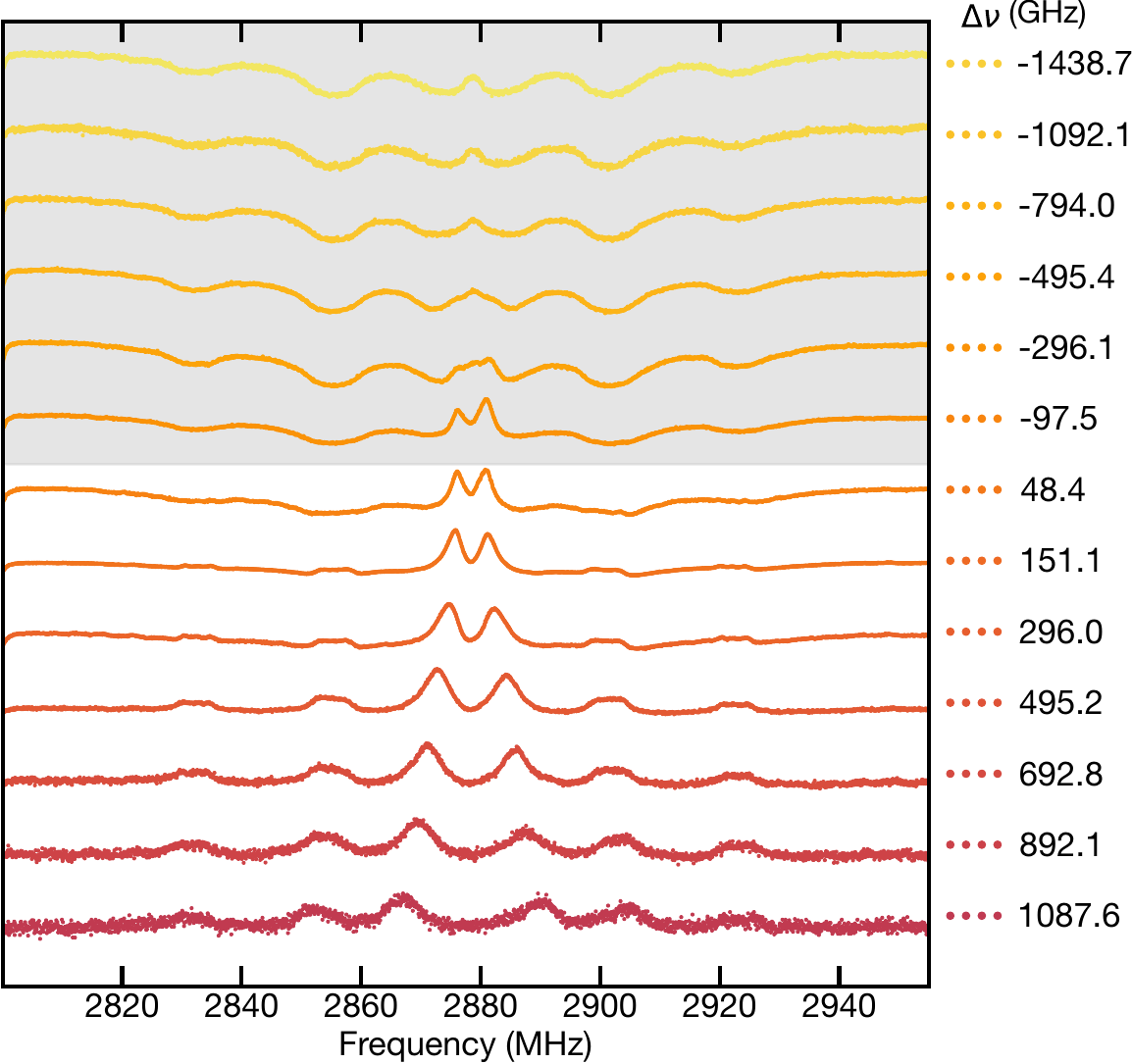}
   \caption{Resonant ODMR spectra at varying optical detuning; shaded background corresponds to detunings above ZPL, while white background corresponds to detunings below ZPL. The spectra were taken with a $20~\si{\gauss}$ magnetic field applied in the plane of the (111)-cut diamond (i.e.~perpendicular to one NV axis). Smaller peaks, split $\gtrsim 20~\si{\mega \hertz}$ from the center frequency, correspond to microwave resonance with other NV groups. Other presentations of these spectra restrict focus to the central group at detunings below ZPL.}
   \label{fig:data-summary}
\end{figure}

Shown in Fig.~\ref{fig:data-summary} is the dependence of the resonant ODMR lineshapes for a wide range optical detunings. 
For drives below ZPL, the positive-contrast peaks are clearly visible and are described quantitatively by our microscopic model (see section \ref{sec:inverted-contrast}).
In contrast, for drives above ZPL, the positive-contrast peaks disappear with increasing detuning.
We conjecture that this disappearance is related to the fact that the resonance condition is most likely to be met by the \emph{upper branch} of the $^3E$ manifold at these detunings, but this branch is itself within the phonon-sideband of the lower branch.
Hence, excited states of the upper branch will have shorter lifetimes than their lower branch counterparts, possibly rendering the linewidth of the associated optical transitions too large for the positive-contrast feature to emerge.

\begin{figure}
   \centering
   \includegraphics[scale=0.45]{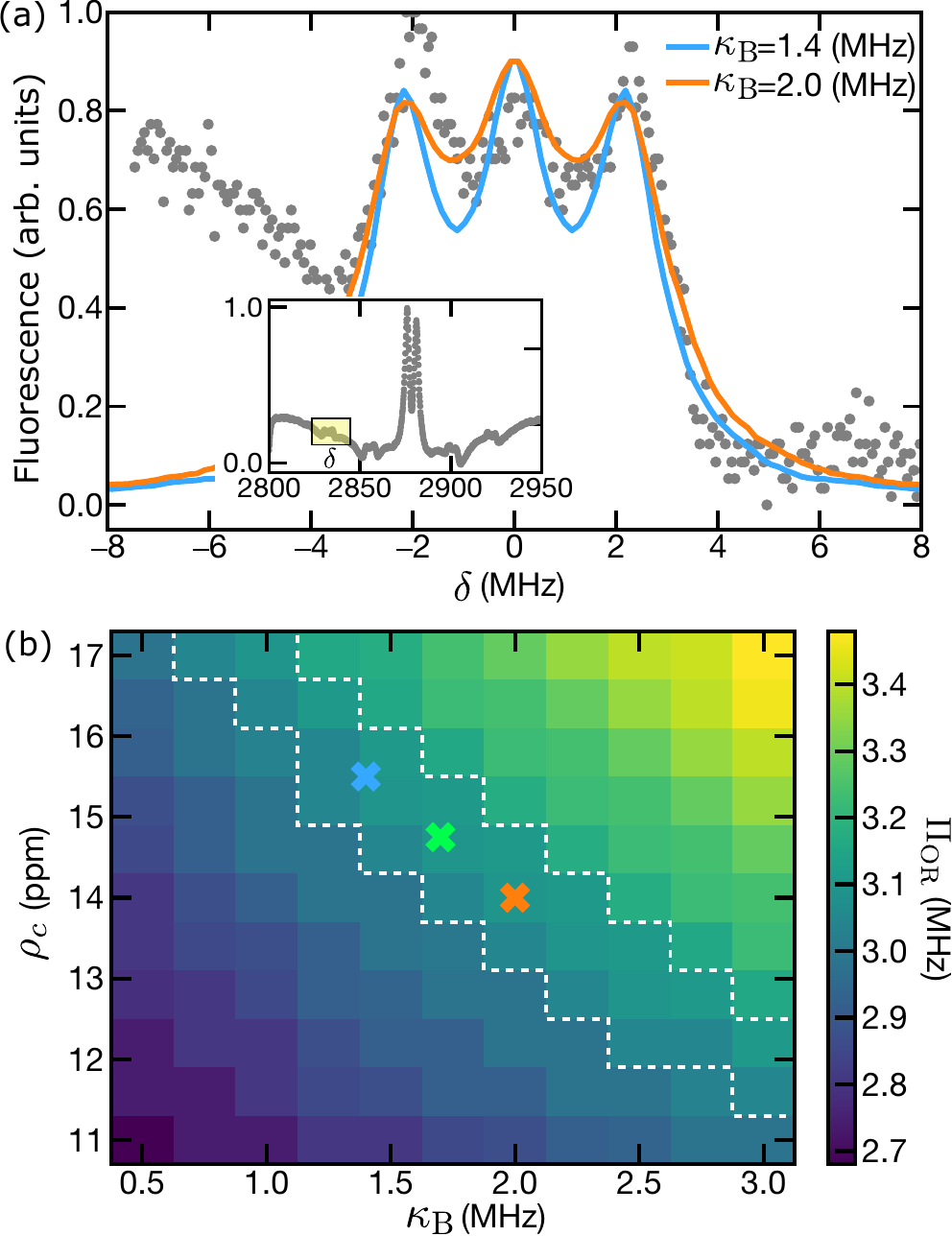}
   \caption{(a) Resonant ODMR spectrum with optical drive detuned $190~\si{\giga \hertz}$ below ZPL and magnetic field applied in the plane of the (111)-cut diamond. We focus on the lineshape of an NV sub-ensemble experiencing a large magnetic field projection along its axis. Solid blue and orange traces are triple-Lorentzian lineshapes with widths $1.4~\si{\mega\hertz}$ and $2.0~\si{\mega\hertz}$ respectively; these are used to constrain the magnetic broadening, $\kappa_\mathrm{B}$. (inset) The full resonant ODMR spectrum. The peaks shown in the main panel are located in highlighted box. (b) Predicted off-resonant ODMR splitting as a function of charge density $\rho$ and $\kappa_\mathrm{B}$. The white-dashed contour indicates the region for which the predicted splitting value is consistent with the room-temperature spectrum [Fig.~1(a), inset]. This region, coupled with the extracted range for $\kappa_\mathrm{B}$, is used to constrain the acceptable values of $\rho$. We extract susceptibilities for three pairs $(\rho, \kappa_\mathrm{B})$, indicated by the colored x-markers, spanning this range.}
   \label{fig:charge-kappa}
\end{figure}

\section{Additional Details of the Resonant ODMR Model} \label{app:model}
In this appendix, we elaborate on four aspects of our microscopic model of resonant ODMR. 
First, we demonstrate that the electric field distribution arising from randomly placed charges can be accurately modeled by the analytic expression equation~\eqref{eq:P(E)}.
Second, we discuss the single-NV primitive lineshape, $\Lambda(\omega;E_\perp)$, used in our analysis. 
Third, we provide the explicit expression for the off-resonant contribution to the resonant ODMR lineshape, $S_\mathrm{OR}$.
Finally, we extend our model to account for the background fluorescence of resonant ODMR, yielding the prediction of Fig.~\ref{fig:3}(b).

\textbf{Electric field distribution:} At its core, our microscopic model proposes that the NV is affected by internal electric fields arising from randomly placed point charges in the diamond lattice.
To determine the electric field distribution, we numerically sampled random spatial configurations for $\sim 100$ charges at a given density $\rho$ and computed their net electric field at an arbitrary spatial point corresponding to the location of the NV. 
Empirically, we found that these Monte Carlo results were well approximated by Eq.~\eqref{eq:P(E)}, especially near the peak of the distribution, which is most relevant for our susceptibility analysis [Fig.~\ref{fig:background}(b)].
Incidentally, Eq.~\eqref{eq:P(E)} corresponds to the electric field distribution owing to the nearest (single) charge at a renormalized density $2 \rho$; however, we consider this a mathematical coincidence.
For our purposes, it is relevant only that Eq.~\eqref{eq:P(E)} provides a convenient analytic expression to approximate the full electric field distribution generated by Monte Carlo simulations

\textbf{Primitive lineshape:} As discussed in the main text, the single-NV ODMR lineshape $\Lambda$ depends on an magnetic broadening parameter $\kappa_\mathrm{B}$, arising from local magnetic fields, and a non-magnetic broadening parameter $\kappa_0$, due to microwave power broadening and strain.
The difference between the two forms of broadening is that magnetic broadening adds in quadrature with the electric field splitting, while non-magnetic broadening is treated as an overall convolution.
We model both forms of broadening as a Lorentzian distribution, where $\kappa$ is the full-width-half-maximum (FWHM).
Finally, we take into account the effective magnetic field, $\mu_B g_e B_\textrm{I} \in \{0, \pm 2.16 \textrm{ MHz}\}$ owing to the three distinct $^{14}$N nuclear states, i.e.~$m_I = 0,\pm 1$.
Altogether, the explicit form for  $\Lambda(\omega;E_\perp)$ is given by
\begin{equation}
\Lambda(\omega;E_\perp) = \int d\omega^\prime \Lambda_\textrm{B}(\omega;E_\perp) \frac {\frac{\kappa_0}{2}} {\pi\left[(\omega-\omega^\prime)^2 + (\frac{\kappa_0}{2})^2\right]} ,
\end{equation} \label{eq:lambda_B}
where
\begin{widetext}
\begin{equation*}
\Lambda_\textrm{B}(\omega;E_\perp)  = \begin{cases}
      0 & \abs{\omega} \leq \chi^\mathrm{g}_\perp E_\perp \\
      \sum_{B_\textrm{I}} \frac{\frac{\kappa_\textrm{B}}{2} \abs{\omega}}{\pi\sqrt{\omega^2-(\chi^\mathrm{g}_\perp E_\perp)^2}\left(\left(\abs{\mu_B g_e B_\textrm{I}}-\sqrt{\omega^2-(\chi^\mathrm{g}_\perp E_\perp)^2}\right)^2+(\frac{\kappa_\textrm{B}}{2})^2\right)} & \abs{\omega} > \chi^\mathrm{g}_\perp E_\perp
   \end{cases} 
\end{equation*}
\end{widetext}
is the lineshape with magnetic broadening alone.

\textbf{Off-resonant contribution:} The expression for the off-resonant contribution to the resonant ODMR spectra is structurally identical to the resonant case.
The essential difference lies in replacing the kernel defining the resonant condition, $D_\textrm{R}$, with a new kernel that quantifies the degree of \emph{off-resonant} driving.
Specifically, the off-resonant kernel, $D_\textrm{OR}$, can take three values: 0 if the optical drive is below both branches, 1 if it is between the two branches, and 2 if it is above both branches; this is because the phonon sidebands of each excited state branch can contribute to the off-resonant cross-section.
Formalizing this physical picture, we obtain,
\begin{align}
    S_\mathrm{OR}(\omega;\Delta \nu) &= \int dE \; P(E) \times \nonumber \\ \int &\sin(\theta) d\theta \Lambda(\omega; \sin(\theta) E)
    D_\mathrm{OR}(\vec E,\Delta \nu), \\
    D_\mathrm{OR}(\vec E,\Delta \nu) & = \left[\Theta(\delta_{U} - \gamma_\mathrm{e}/2) + \Theta(\delta_{L} - \gamma_\mathrm{e}/2) \right].
\end{align}

\textbf{Total fluorescence:} The total fluorescence is determined by the fraction of resonant and off-resonant configurations.
These fractions are given by
\begin{align}
    F_\mathrm{R}(\Delta \nu) &= \frac{1}{\mathcal{N}} \int dE\;P(E) \times \nonumber \\ & \int \sin(\theta) d\theta
    D_{R}(\vec E,\Delta \nu) \\
    F_\mathrm{OR}(\Delta \nu) &= \frac{1}{\mathcal{N}} \int dE \; P(E) \times \nonumber \\ & \int \sin(\theta) d\theta
    D_{OR}(\vec E, \Delta \nu)\ ,
\end{align}
where $\mathcal{N}$ is the total number of configurations.
The total fluorescence is a weighted sum of these two contributions:
\begin{equation} \label{eq:R}
    R(\Delta \nu) \propto \epsilon_\mathrm{R} F_\mathrm{R}(\Delta \nu)+F_\mathrm{OR}(\Delta \nu)\ ,
\end{equation}
where $\epsilon_\textrm{R}$ is the enhancement factor of the resonant mechanism.
From single NV experiments, we estimate $\epsilon_\textrm{R} \approx 10^5$ \cite{robledo_high-fidelity_2011}.
Up to overall rescaling, we can then calculate the predicted fluorescence as a function of detuning; this exhibits good agreement with the background fluorescence as shown in Fig.~\ref{fig:3}(b).

\section{Estimating Susceptibilities} \label{app:susceps}
As discussed in the main text, we extract the excited-state electric field susceptibilities by fitting our model to the measured splitting of the positive-contrast peak, $\Pi_\perp$, as a function optical detuning, $\Delta \nu$.
Here, we provide additional details on this procedure, including error estimation and the determination of model parameters, i.e.~the charge density and broadening parameters.

\begin{figure}
   \centering
   \includegraphics[scale=0.5]{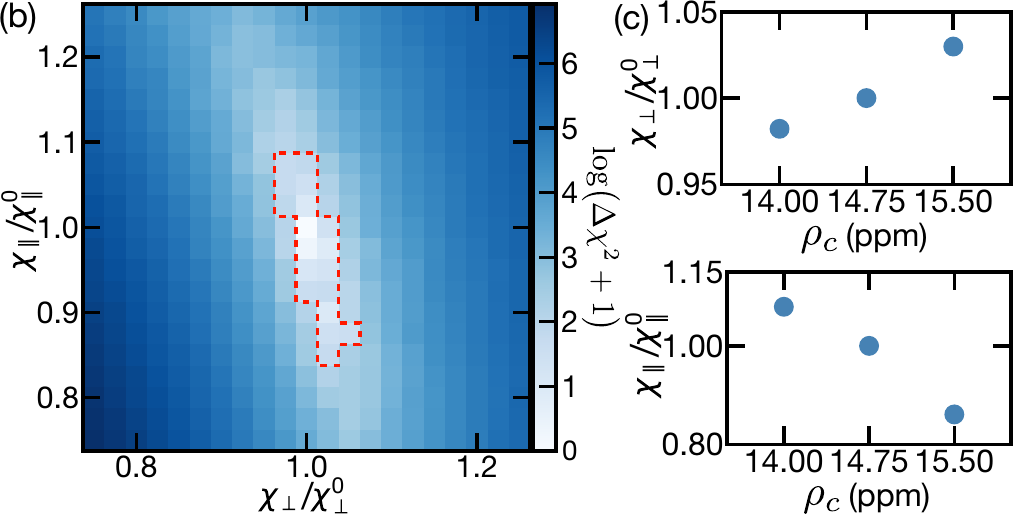}
   \caption{(a) Experimental $\Pi_\perp$ (dark blue) and model (gray) as functions of $\Delta \nu$. The fit yields $\chi_\nu^2=0.87$ (b) $\Delta \chi^2$ as a function of $\chi^\mathrm{e}_\parallel$ and $\chi^\mathrm{e}_\perp$. The red-dashed contour denotes the $2 \sigma$ confidence region. (c) $\chi^\mathrm{e}_\perp$ and $\chi^\mathrm{e}_\parallel$ as functions of $\rho$. This quantifies the main source of systematic error in our analysis. Errors are relative to $\{\chi^0_\perp,\chi^0_\parallel\}  = \{1.43,0.68\} ~\si{\mega\hertz/ (\volt/\centi \meter)}$}
   \label{fig:susc-errors}
\end{figure}

To begin, we determine $\Pi_\perp$ as a function of detuning from the experimentally measured ODMR spectra [Fig.~2(a)].
In particular, we identify the frequency of the local maximum, $\omega_{\pm}$, associated with each positive-contrast peak and compute $\Pi_\perp = \frac 1 2 (\omega_{+} - \omega_{-})$.
The uncertainty on these estimates arises from shot noise in the resonant ODMR spectra, which causes the frequency of maximum florescence to vary between successive measurements.
To determine this uncertainty, we perform a Monte Carlo simulation of Lorentzian lineshapes with Gaussian noise, whose strength is determined from the experimental data, and sample the frequency of local maximum; this yields the error bars shown in Fig.~2(b) and \ref{fig:susc-errors}(a).

We next determine the three parameters required in our resonant ODMR model (see section \ref{subsec:inverted-contrast-model}) other than the susceptibilities through the following independent calibration steps:
\begin{enumerate}
    \item Magnetic broadening, $\kappa_\mathrm{B}$: We measure the resonant ODMR spectrum of an NV sub-ensemble with a significant magnetic field projection along its axis [Fig~\ref{fig:charge-kappa}(a)].
    Since this magnetic field suppresses electric field noise, the dominant source of remaining noise is due to inhomogeneous magnetic fields.
    Fitting this spectrum to three Lorentzians yields an magnetic linewidth $\kappa_\mathrm{B} = 1.7 \pm 0.3~\si{\mega \hertz}$ [Fig.~\ref{fig:charge-kappa}(a)].
    \item Charge density, $\rho$: We measure a room-temperature, off-resonant ODMR spectrum without a bias magnetic field.
    The characteristic splitting observed in this spectrum is fit to our model of randomly placed charges, leading to a charge-density estimate of $\rho = 15 \pm 2$ ppm.
    \item Non-magnetic broadening, $\kappa_0$: We perform a Lorentzian fit to the positive-contrast features of an ODMR spectrum measured with optical excitation near the zero-phonon-line ($\Delta \nu \lesssim 1~\si{\giga \hertz}$).
    This spectrum is chosen because it has minimal broadening due to electric fields.
    We subtract $\kappa_\mathrm{B}$ from the extracted linewidth and assume the remaining broadening arises from non-magnetic sources (e.g.~microwave power broadening); this yields $\kappa_0 \approx 1~\si{\mega \hertz}$.
    We note that this parameter has only a minor effect on the susceptibility estimates.
\end{enumerate}

Finally, we extract the susceptibility parameters by fitting our model to the empirical values for $\Pi_\perp$ as a function of $\Delta \nu$.
In particular, we calculate the least-square error of the data compared to the predicted splittings from our resonant ODMR model, with $\chi^\mathrm{e}_\perp$ and $\chi^\mathrm{e}_\parallel$ as the only free parameters [Fig.~\ref{fig:susc-errors}(a)]\footnote{In the vicinity of the positive-contrast peaks, the off-resonant configurations only contribute a flat background, and can hence be neglected in our fitting procedure.
This eliminates $\epsilon_C$ and off-resonant linewidths as free parameters and thus simplifies the model we use to extract the susceptibilities.
To check this approximation, we also extract the susceptibilities using the $\epsilon_C$ and off-resonant linewidths extracted from the $5~\si{\kelvin}$ spectrum (Table~\ref{tab:temp}) and find that they are within statistical error of those determined from our simpler estimation procedure.}.
We find the $\chi^2$-error is minimized at $\{\chi^\textrm{e}_\perp,\chi^\textrm{e}_\parallel\}  = \{1.43,0.68\} ~\si{\mega\hertz/ (\volt/\centi \meter)}$ with a reduced-$\chi^2$ value of $\chi^2_\nu = 0.87$ (with $15$ observations and $2$ fit parameters).
By linearizing our model around the fitted values, we determine the $2\sigma$ confidence region of the susceptibility estimates [Fig.~\ref{fig:susc-errors}(b)] and estimate uncertainties of $\sim 5\%$ for $\chi^\mathrm{e}_\perp$, and $\sim 15\%$ for $\chi^\mathrm{e}_\parallel$.
We also estimate systematic errors by repeating the analysis with the values $\rho$ and $\kappa_\mathrm{B}$ indicated in Fig~\ref{fig:charge-kappa}(b).
This is shown in Fig.~\ref{fig:susc-errors}(b) and leads to a systematic error of $\sim 5\%$ for $\chi^\mathrm{e}_\perp$, and $\sim 15\%$ for $\chi^\mathrm{e}_\parallel$.
Summing in quadrature, we have a total error estimate of $\sim 7\%$ for $\chi^\mathrm{e}_\perp$ and $\sim 21\%$ for $\chi^\mathrm{e}_\parallel$.

A few additional remarks are in order.
First, we note that our procedure, by focusing on the splitting of the ODMR spectra, depends primarily on the magnitude of the most-probable electric field (i.e.~$E_0$) and not on the details of the full distribution.
Second, in principle $\rho$ may depend on the temperature, optical excitation frequency, and excitation power, which would invalidate our assumption that $\rho$ can be determined from off-resonant room-temperature ODMR \cite{PhysRevLett.107.266403, acosta_dynamic_2012, manson_nv-_2018}.
However, even if we relax this assumption, the optical transition linewidth provides an additional, independent constraint on the charge-density and susceptibilities.
By demanding that $(\rho, \chi^\mathrm{e}_\parallel, \chi^\mathrm{e}_\perp)$ simultaneously recover $\Pi_\perp(\Delta \nu)$ and the excited state linewidth, we find $\rho = 15^{+7}_{-3}$, where the super- (sub-) script indicates the upper (lower) bound.
The systematic errors on the extracted susceptibilities concordantly increase to $\sim 10\%$ and $\sim 30\%$ for $\chi^\mathrm{m}_\perp$ and $\chi^\mathrm{m}_\parallel$ respectively.
Therefore, the essentials of our analysis and conclusions do not depend on an assumption of consistent $\rho$ (although our observations support this conclusion for ensemble measurements).

\section{Resonant ODMR Linewidth} \label{app:linewidth}

\begin{figure}
   \centering
   \includegraphics[scale=0.5]{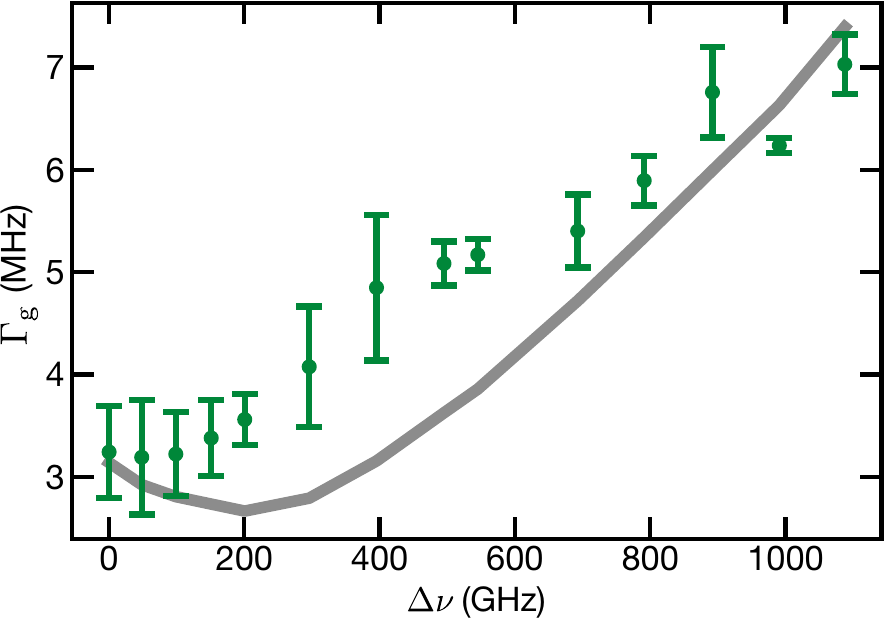}
   \caption{Resonant ODMR linewidth $\Gamma_\mathrm{g}$ as a function optical detuning. Error bars reflect the difference in the FWHM of Lorentzian fits of the left and right peaks of experimental data. The same analysis applied to the spectra generated by our model yields the gray curve. Although the model accounts for the general trend of increasing $\Gamma$ at large detuning, there are clear qualitative differences between the experiment and theory. Most notably, the experiment is broader at moderate detunings ($200-600~\si{\giga \hertz}$) than the model would suggest. This could be because the true electric field distribution decays more slowly than the random charge model predicts. The effects of strain may also partially account for the discrepancy.}
   \label{fig:gamma-v-detuning}
\end{figure}

The positive-contrast features shown in Fig.~\ref{fig:2}(a) exhibit not only a splitting $\Pi_\perp$ which depends on the optical detuning, but also a linewidth $\Gamma_\mathrm{g}$ which systematically increases with detuning (Fig.~\ref{fig:gamma-v-detuning}).
Qualitatively, this effect can be understood as arising from the \emph{tail} of the electric field distribution: If the electric field distribution decays very slowly, $\Gamma_\mathrm{g}$ will be large since many nearly equal probability electric-field spheres will intersect the resonant cone at different values of $E_\perp$.

While our microscopic model accounts for this general trend, it does not accurately predict the precise form of $\Gamma_\mathrm{g}$ vs.~$\Delta \nu$ (Fig.~\ref{fig:gamma-v-detuning}).
Interestingly, this discrepancy suggests that our microscopic model is missing subtle aspects of the electric field distribution at large strengths / short distances.
More formally, the trend that $\Gamma_\mathrm{g}$ increases with $\Delta \nu$ (Fig.~\ref{fig:gamma-v-detuning}) indicates the relative decay rate of the electric-field distribution \emph{decreases} at larger values of $E$.
This is characteristic of a polynomially decaying tail: if $P(E) \sim 1/(E/E_0)^q$ then $\frac{d(\log(P(E)))}{dE} = -q/E$, so the tail decays more slowly at larger $E$ resulting in larger $\Gamma_\mathrm{g}$.
Indeed, one possible direction for future research is to quantitatively extract $\frac{d(\log(P(E))}{dE}$ from $\Gamma_\mathrm{g}(\Delta \nu)$ at large $\Delta \nu$.
This could yield insight into the underlying short-range physics controlling the tail of the electric field distribution, such as whether charges are more likely to be localized in the vicinity of the NV center.

%

\section{Temperature Dependent Spectra} \label{app:temp}
Our procedure for fitting the temperature-dependent ODMR spectra, shown in Fig.~1(a), relies on the same resonant ODMR model, susceptibility parameters, and charge density as in the previous sections.
In addition, we find it necessary to vary (i) the ODMR contrast ratio and (ii) broadening parameters at each temperature.
Here, we discuss the physical origin of the temperature dependence of these parameters.

\textbf{Contrast:} We attribute the change in contrast to the the fact that increasing temperature broadens the optical transition linewidth, which in turn reduces the density of resonant configurations at a given optical detuning, effectively reducing $\epsilon_C$.
Indeed, we find qualitatively that $\epsilon_C$ decreases with temperature, though we do not attempt to develop a quantitative model for it.

\textbf{Broadening:}
Experimentally, we observe that off-resonant dips in the ODMR spectrum exhibit larger broadening than the positive-contrast peaks (see Fig.\,1(a) in main text).
We conjecture two possible causes of this additional broadening.
First, there may be different degrees of \emph{light-narrowing} in the resonant and off-resonant configurations.
Light-narrowing arises because resonantly driven transitions experience a greater rate of optical pumping than transitions to the phonon-sideband, and hence have a different effective linewidth.
In this setting, a greater rate of optical pumping actually \emph{reduces} the effective linewidth \cite{jensen_light_2013}.
Second, power broadening itself may not be well described by a convolution with a Lorentzian; indeed we find that this is the case even for \emph{off-resonant} ODMR spectra under higher microwave power.
Intuitively, large electric fields alter the matrix elements between ground-state sublevels, causing the degree of power-broadening to be electric field strength dependent.
To account for this, we treat the off-resonant magnetic broadening parameter, $\kappa_\mathrm{B}^\mathrm{OR}$, as a free parameter and find that its best-fit value is comparable to the non-magnetic broadening.
We emphasize that this is a convenient way to account for the dependence of power broadening on electric field strength and does not constitute a meaningful estimate of the true broadening due to magnetic fields.

We note that both of these issues are artifacts of working in a microwave power-broadened regime, which is the case for the present temperature-dependent data.
For extraction of susceptibilities, ODMR measurements were performed with $100~\times$ lower microwave power, where these effects are suppressed.

\renewcommand{\arraystretch}{1.5}
\begin{table}
\centering
\begin{tabular}{c|c|c|c|c|c}
\hline \hline
 \multirow{2}{*}{} $\textrm{Temp.}$ & $\kappa^\mathrm{R}_0$ & $\kappa^\mathrm{R}_\mathrm{B}$ & $\kappa^\mathrm{OR}_0$ &
 $\kappa^\mathrm{OR}_\mathrm{B}$ &
 $\epsilon_C$ \\
 \multirow{2}{*}{} $\left[ \si{\kelvin} \right]$ & $\left[ \si{\mega \hertz} \right]$ & $\left[ \si{\mega \hertz} \right]$ & $\left[ \si{\mega \hertz} \right]$ &
 $\left[ \si{\mega \hertz} \right]$ &
 $\left[ 1 \right]$ \\
\hline
 $5$ & 2.0 & 4.0 & 20.0 & 27.0 & $10^4$ \\
 \hline
 $40$ & 2.0 & 4.0 & 16.0 & 16.0 & $4 \cdot 10^3$\\
 \hline
 $55$ & 2.0 & 4.0 & 12.0 & 15.0 & $-1.7 \cdot 10^3$\\
 \hline
 $100$ & 2.0 & 4.0 & 9.0 & 8.0 & $-1.7 \cdot 10^3$\\
 \hline
 \hline
\end{tabular}
\caption{Summary of parameters used to fit the temperature dependent ODMR spectra [Fig.\,1(a)]. The linewidth parameters for resonant configurations are roughly consistent with the more carefully estimated parameters used for susceptibility extraction; the off-resonant linewidth parameters, however, are significantly altered by power-broadening, as we discuss in Section\,\ref{app:temp}. In particular, $\kappa^\mathrm{OR}_\mathrm{B}$ should not be interpreted as an accurate estimate of the magnetic broadening; instead, we regard it as a phenomenological parameter that controls the amount of broadening which ``adds in quadrature" to the electric field [see equation~\eqref{eq:lambda_B}]. The value of $\epsilon_C$ at $5~\si{\kelvin}$ is roughly consistent with an estimate based on $\epsilon_R$. Qualitatively, $\epsilon_C$ decreases with increasing temperature because the density of resonant states decreases.}
\label{tab:temp}
\end{table}

\section{Sensitivity Estimates} \label{app:electrometry}
Here, we provide additional details for estimating the sensitivity of our resonant electric-field sensing protocol.
The estimates are calibrated based on the sample measured in this work, and then extrapolated to other densities using scaling arguments.

Let us begin by recalling that the sensitivity for an electric-field dependent count rate $R(\delta E)$ is given by
\begin{align}
    \frac{1}{\eta} = \frac{dR}{d(\delta E)}\biggm\lvert_{\delta E = 0} \frac{1}{\sqrt{R(0)}}, \\
\end{align}
or equivalently,
\begin{equation}
    \frac 1 \eta = \frac{d \log(R)}{d(\delta E)}\biggm\lvert_{\delta E = 0} \cdot \sqrt{R(0)}\ .
 \end{equation}
%
%
In our sensing protocol, the fluorescence rate is determined by two effects: (i) the shift of the positive-contrast peaks, and (ii) the change in overall fluorescence.
These effects contribute independently to the total signal, such that $R(\delta E) = R_\mathrm{F}(\delta E) \cdot R_\Pi(\delta E)$, where $R_\mathrm{F}$ and $R_\Pi$ capture the dependence of fluorescence on optical detuning and microwave frequency, respectively.
The total sensitivity is then
\begin{equation}
\frac 1 \eta = \left[\frac{d \log(R_\Pi)}{d (\delta E)} + \frac{d \log(R_\mathrm{F})}{d (\delta E)}\right]_{\delta E = 0} \cdot \sqrt{R(0)}\ ,
\end{equation}
which leads to the equation stated in the main text:
\begin{equation}
\frac 1 \eta = \frac 1 {\eta_\Pi} + \frac 1 {\eta_\mathrm{F}}\ .
\end{equation}
For simplicity, we define $R \equiv R(0)$.
The sensitivities can be decomposed as
\begin{align} \label{eqn:sensing-formulas}
\eta_{\Pi} &= P_\Pi\frac{\Gamma_{\textrm{g}}}{\chi_\textrm{eff} C_0 C_\mathrm{r}}\cdot \frac 1 {\sqrt R}\ , \\
\eta_{\textrm{F}} &= P_{\textrm{F}} \frac{\Gamma_{\textrm{e}}}{\chi^{\textrm{e}}_{\parallel} C_\mathrm{r}}\cdot \frac 1 {\sqrt R}\ ,
\end{align}
where $\Gamma~(P)$ is the linewidth (lineshape factor) associated with the ground and excited states, $R$ is the count rate, $C_0$ is the maximum CW contrast of the resonant ODMR peaks, $C_\textrm{r}$ is the ratio of photons resulting from resonant fluorescence to total fluorescence, and $\{\chi_\textrm{eff}, \chi^\textrm{e}_\parallel\}$ are the effective ground- and excited-state susceptibilities, respectively.

For our estimates, we assume the ground state ODMR lineshape is Lorentzian, such that $P_\Pi \approx 0.77$, and we use the excited-state susceptibility determined in our work: $\chi^\mathrm{e}_\parallel = 0.7 ~\si{\mega\hertz/ (\volt/\centi \meter)}$; moreover, we determine $\chi_\mathrm{eff} \approx 0.41$ and $P_\mathrm{F} \approx 0.39$ from our model [Fig.~\ref{fig:sensing-ests}(a)] and from experimental data [Fig.~\ref{fig:3}(b)], respectively.
The remaining parameters are discussed below.

\subsection{Parameter calibration}\label{subapp:param-cal}
\textbf{Linewidths:}
We model the ODMR and optical transition linewidths as containing both an intrinsic broadening and an electric-field induced broadening:
\begin{equation}
    \Gamma_\mathrm{g,e} = \Gamma^0_\mathrm{g,e}(\bar{\rho}) + \Gamma^E_\mathrm{g,e} \cdot \bar{\rho}^{2/3}\ ,
\end{equation}
where $\bar{\rho}$ is a normalized NV density, $\bar{\rho}=\rho_\mathrm{NV}/\rho^0_\mathrm{NV}$.
The ground state intrinsic broadening, $\Gamma^0_\mathrm{g}(\bar{\rho})$, is typically dominated by paramagnetic impurities and is modeled by \cite{edmonds_generation_2020}
\begin{equation}
        \Gamma^0_\mathrm{g} = (A_{^{13} \mathrm{C}} [^{13}\mathrm{C}]) + A_{\mathrm{N}^0} [\mathrm{N}^0])^{-1} \cdot \frac 1 \pi , 
\end{equation}
where $[^{13}\mathrm{C}], [\mathrm{N}^0]$ represent the concentrations of $^{13} \mathrm{C}$ and uncharged substitutional nitrogen defects, respectively.
From \cite{edmonds_generation_2020} we obtain $A_{^{13} \mathrm{C}} \approx 0.1~\si{\milli \second}^{-1} \mathrm{ppm}^{-1}$ and $A_{\mathrm{N}^0} \approx 101~\si{\milli \second}^{-1} \mathrm{ppm}^{-1}$.
For impurity concentrations, we assume a natural abundance of $^{13}\mathrm{C}$ ($1.1 \%$ or $11000$ ppm) and that $[\mathrm{N}^0] = 0.3 \cdot [\mathrm{NV}^{-1}]$ --- i.e., a $30 \%$ conversion ratio, which is among the best typically observed \cite{barry_sensitivity_2020}.
Since the intrinsic excited-state broadening is less well-understood, and certainly not limited by paramagnetic impurities, we adopt the simple density-independent model $\Gamma^0_\mathrm{e} \approx 10~\si{\giga \hertz}$ \cite{batalov_temporal_2008, PhysRevLett.110.213605}.
Finally, we calibrate the charge-induced linewidths against our sample ($\rho^0_\mathrm{NV} \approx 8$ ppm), yielding $\Gamma^E_\mathrm{g} \approx 3.7$ $\si{\mega \hertz}$ and $\Gamma^E_\mathrm{e} \approx 10^6$ $\si{\mega \hertz}$.
We note that here, and throughout our sensitivity estimates, we will assume that the charge environment consists primarily of other uniformly distributed NVs and charge donors, such that $\rho = 2 \rho_\mathrm{NV}$ (see subsection \ref{subsec:inverted-contrast-model}).

\begin{figure} \centering
   \includegraphics[scale=0.5]{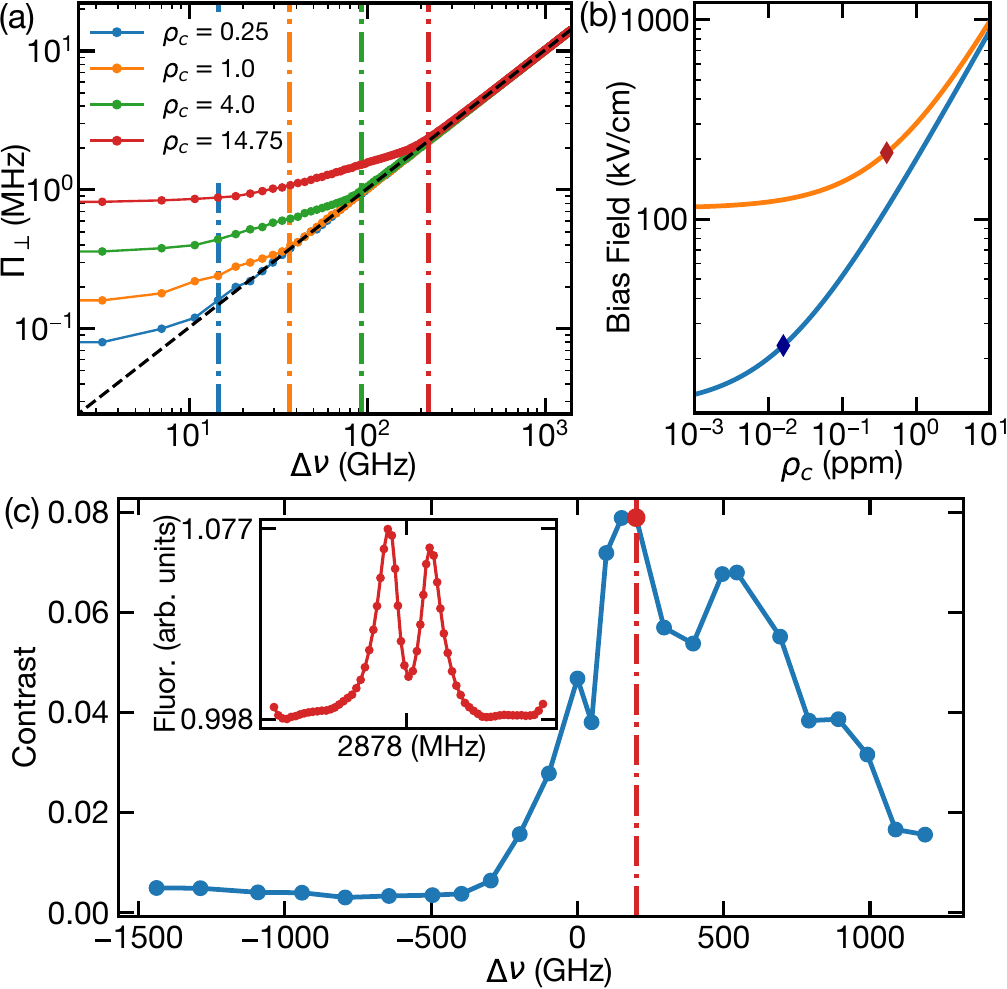}
   \caption{(a) $\Pi_\perp$ as a function of $\Delta \nu$ for various $\rho$. Dash-dot vertical lines indicate the optimal operating $\Delta \nu$ for our sensing proposal. In particular, we choose the smallest $\Delta \nu$ for which $\Pi_\perp$ depends linearly on $\Delta \nu$. This maximizes $\chi_\mathrm{eff}$ and the resonant fluorescence. (b) Fluorescence as a function of $\Delta \nu$ from experiment (brown) and theory (solid blue). The dash-dot vertical line indicates the optimal operating $\Delta \nu$  for our sample. We estimate $\Gamma_\mathrm{e} \approx 1~\si{\tera \hertz}$, ignoring asymmetry (dashed orange lines). (c) Contrast of resonant ODMR (with a $20~\si{\gauss}$ applied magnetic field perpend icular to the NV axis) as a function of $\Delta \nu$. The red marker and dash-dot line indicates the contrast at the optimal operating detuning. Since magnetically split groups provide additional background for the central peak, the maximum CW ODMR contrast $C_0$ is a factor $8/3$ larger than what is observed (see Section~\ref{subapp:param-cal}) \cite{barson_nanomechanical_2017}. (inset) Experimental resonant ODMR at the optimal operating $\Delta \nu$. (d) Electric field bias required for our sensing procedure as a function of $\rho$ assuming $\kappa_\mathrm{e}^0=10~\si{\giga \hertz}$ (blue) and $\kappa_\mathrm{e}^0=100~\si{\giga \hertz}$ (orange). The blue and red diamonds mark the bias field required at optimal NV densities.}
   \label{fig:sensing-ests}
\end{figure}

\textbf{Fluorescence rate:}
The overall fluorescence rate contains contributions from the resonant and off-resonant configurations.
The former is proportional to the density of resonant configurations, and therefore is inversely proportional to the optical transition linewidth.
Thus, the fluorescence from resonant configurations $R_R$ can be modeled as
\begin{align}\label{eqn:r}
    R_R = r R_0 ,\quad r = \frac{\Gamma^\mathrm{ref}}{\Gamma^0_\mathrm{e} + \Gamma^E_\mathrm{e} \bar{\rho}^{2/3}}\ ,
\end{align}
where $R_0$ is the fluorescence rate for a single, off-resonantly driven NV orientation, $r$ is the fluorescence enhancement factor for resonant configurations, and $\kappa^\textrm{ref}$ is a density-independent prefactor.
To determine $\Gamma^\textrm{ref}$, we compare the resonant fluorescence at the optimal detuning to the off-resonant fluorescence for detunings far above the ZPL [Fig.~\ref{fig:sensing-ests}(b)].
This yields $r \approx 2$, corresponding to $\Gamma^\mathrm{ref} \approx 2 \cdot 10^6~\si{\mega \hertz}$; for an optimal sample, we estimate that $r$ can reach $r \approx 100$ at $\sim 10$ ppb.

To determine the overall count rate, we take into account the fact that our sensing protocol includes signals from one resonant NV orientation and three off-resonant orientations.
This is because the bias electric field required to lift the excited-state degeneracy pushes the excited state of three orientations below the excited state of the target orientation, so they are excited by the off-resonant mechanism.
In particular, assuming a (111)-cut diamond, the fluorescence rate due to off-resonant orientations is $R_\textrm{OR} = 5/3~R_0$~\footnote{The factor of 5/3 arises because the three off-resonant groups are not perpendicular to the laser polarization \cite{barson_nanomechanical_2017}.}.
In combination, the total count rate is
\begin{align}
    R = R_0(r+5/3)\ .
\end{align}
We note that $R_0$ depends on the specific optical setup (e.g. illumination volume, laser power, and collection efficiency).
We determine this constant for a similar setup as described in \cite{chen_high-sensitivity_2017}
\footnote{In particular, we rescale their reported count rate by the ratio of our sample's NV density to their sample's density and correct for a difference in laser polarization.
We also divide by $8$, since only one NV crystallographic orientation will be used in the sensing protocol.}.
The final count rate for our sample and for an optimal sample are reported in Table \ref{tab:sensitivity_param}.

Finally, for the room-temperature sensitivity estimate (Fig.~3), we additionally adjust $R$ to account for the decrease in the Debye-Waller factor with temperature.
We estimate the Debye-Waller factor decreases by $35 \%$ at $300 \si{\kelvin}$ compared to $5 \si{\kelvin}$, degrading the sensitivity by $\sim 50 \%$ \cite{plakhotnik_all-optical_2014,kehayias_infrared_2013}.

\renewcommand{\arraystretch}{1.5}
\begin{table*} \label{tab:electrometry}
\centering
\begin{tabular}{l|c|c|c|c|c|c|c|c|c|c}
 \hline \hline
  \multicolumn{2}{c|}{DC Electrometry} & \makecell{$\rho_\mathrm{NV}$} & T& $\Delta \nu_{\rm op}$& \makecell{$\Gamma$} & \makecell{$P$} & \makecell{$\chi$} & \makecell{$R$} & \makecell{$C$} & \makecell{$\eta$}\\
  \multicolumn{2}{c|}{Method} & \makecell{(ppm)} & \makecell{(\si{\kelvin})} & \makecell{(\si{\GHz})}& \makecell{(\si{\mega\hertz})} & & \makecell{(\si{\hertz~\centi\meter/\volt})}& \makecell{(counts/\si{\second})} & & \makecell{(\si{\volt/\centi\meter \sqrt{\hertz}})}\\ \hline \hline
 \multicolumn{2}{l|}{Single NV - G.S. \cite{dolde2011electric}} & - &  $\lesssim 300$ & - & 0.05 & 0.77 & 17 &  100 & 0.3 & 760\\
 \multicolumn{2}{l|}{Single NV - E.S. \cite{yiwen-chu-thesis}} & - & $\lesssim 4$ & - & 13 & 0.77 & $7.0 \cdot 10^5$  &  2500  & 1 & 0.29\\ \hline
 \multicolumn{2}{l|}{EIT Ensemble \cite{PhysRevLett.110.213605}} & $\sim 0.03$ & $\lesssim 30$ & - & 1.0 & 0.77 & 9.8 & $3.2\cdot10^{14}$  & 0.022 & 0.20\\ \hline
 \multicolumn{2}{l|}{Off-res. Ensemble \cite{chen_high-sensitivity_2017} (optimized)} & $\sim 1$ & $\lesssim 300$ & - & 1.0 & 0.77 & 17 & $5.0\cdot10^{14}$  & 0.02 & 0.10\\ \hline \hline
 \multirow{2}{*}{Res. Ensemble} & $F$ & & $\lesssim 200$ &  & $10^6$ & 0.39 & $7.0 \cdot 10^5$ & $2.4\cdot10^{15}$ & 0.54 & 0.021\\
 \multirow{2}{*}{(Our sample)} & $\Pi_\perp$ & $\sim 8$ & $\lesssim 45$ & 200 & 3.9 & 0.77 & 6.97 & $2.4\cdot10^{15}$ & 0.11 & 0.077\\
 &  Total &  & $\lesssim 45$ &  & - & - & - & - & - & 0.017\\ \hline
 \multirow{2}{*}{Res. Ensemble} & $F$ & & $\lesssim 55$ &  & $2.1 \cdot 10^4$ & 0.39 & $7.0 \cdot 10^5$ & $6.9\cdot10^{13}$ & 0.98 & 0.0014\\
 \multirow{2}{*}{(Low-density sample)} & $\Pi_\perp$ & $\sim 0.01$ & $\lesssim 45$ & 2.3 & 0.25 & 0.77 & 6.97 & $6.9\cdot10^{13}$ & 0.21 & 0.016\\
 &  Total &  & $\lesssim 45$ &  & - & - & - & - & - & 0.0013\\ \hline
 \multirow{2}{*}{Res. Ensemble} & $F$ & & $\lesssim 55$ &  & $1.4 \cdot 10^4$ & 0.39 & $7.0 \cdot 10^5$ & $3.1\cdot10^{13}$ & 0.98 & 0.0015\\
 \multirow{2}{*}{([$^{12}\mathrm{C}$] = 99.995\% purified)} & $\Pi_\perp$ & $\sim 0.002$ & $\lesssim 45$ & 0.8 & 0.017 & 0.77 & 6.97 & $3.1\cdot10^{13}$ & 0.21 & 0.0016\\ 
 &  Total &  & $\lesssim 45$ &  & - & - & - & - & - & $8\cdot 10^{-4}$\\
 \hline \hline
\end{tabular}
\caption{Summary of DC NV electrometry protocols and their associated sensitivities, assuming an illumination volume of $\sim 0.1~\si{\milli \meter}^3$ (for ensemble techniques). For our protocol (lower section), we provide both the sensitivity owing to the peak shift ($\eta_\Pi$) and the sensitivity owing to the change in overall fluorescence ($\eta_\mathrm{F}$). All sensitivities are given by $\eta = P \frac{\Gamma}{C \chi \sqrt R}$, where $P$ is a lineshape dependent prefactor, $C$ is the contrast (i.e.~$C=C_0C_\textrm{r}$ for $\eta_\Pi$, and $C=C_\textrm{r}$ for $\eta_\mathrm{F}$), $\chi$ is the relevant susceptibility, and $\Gamma$ is the relevant linewidth (i.e.~$\Gamma=\Gamma_\textrm{g}$ for $\eta_\Pi$, and $\Gamma=\Gamma_\textrm{e}$ for $\eta_\mathrm{F}$). The $^{12}\mathrm{C}$ concentration assumed for an isotopically purified sample is estimated from \cite{edmonds_generation_2020}. For reference \cite{chen_high-sensitivity_2017}, we report the sensitivity that would obtained at optimal NV densities (by rescaling the linewidth and counts according to our scaling formulae). The parameter $\Delta \nu$ refers to the optimal optical detuning below the ZPL for our electrometry protocol [Fig.~\ref{fig:sensing-ests}(a)]}
\label{tab:sensitivity_param}
\end{table*}

\textbf{Contrast:}
We first estimate the maximum CW contrast of the positive-contrast peaks, $C_0 \approx 0.21$, from the experimental data [Fig.~\ref{fig:sensing-ests}(c)]~\footnote{For a (111)-cut diamond, $C_0 = 8/3~C_\mathrm{exp}$, where $C_\mathrm{exp}$ is the experimentally observed contrast. This factor is also related to the laser polarization \cite{barson_nanomechanical_2017}}.
However, of the total counts, only the resonant fraction, $C_\mathrm{r}$, contributes to sensing.
Therefore, the actual contrast of the resonant ODMR peaks is $C = C_0 C_\textrm{r}$; meanwhile, for sensing via direct variation in fluorescence ($\eta_\mathrm{F}$), the relevant contrast is $C_\mathrm{r}$.
For a (111)-cut sample, we have
\begin{align}
    C_\textrm{r} = \frac{r}{(r+5/3)}\ ,
\end{align}
where $r$ is the resonant enhancement factor defined in \eqref{eqn:r}.

\textbf{Bias electric field:} In addition to the sensitivity estimates provided in the manuscript, we  estimate the bias field required to spectrally isolate the crystallographic orientation used in our sensing protocol [Fig.~\ref{fig:sensing-ests}(d)].
Given a bias field parallel to one of the NV groups, three other groups will actually be shifted by the electric field \textit{below} the target group since $\chi^\mathrm{e}_\perp \approx 2 \chi^\mathrm{e}_\parallel$.
Therefore, we determine this bias field by demanding the lowest-energy NV group parallel to the bias field is at least $\Gamma_\mathrm{e}/2$ above the lowest three NV groups.

\subsection{Scaling with density}
Here, we determine how sensitivity degrades in the high-density regime.
We consider an ideal limit for which there is minimal charge broadening, i.e.~$\rho \equiv \rho = 2\rho_\mathrm{NV}$ and no density-dependent broadening other than internal electric-fields.
As discussed above, in the charge-dominated regime the ODMR and optical transition linewidths ($\Gamma_\mathrm{g}, \Gamma_\mathrm{e}$) are proportional to the average electric field strength, which scales as $\rho_\textrm{NV}^{2/3}$.
In this regime, we also have $r \propto \rho^{-2/3}$ and $r \ll 5/3$, implying $C_\mathrm{r} \propto \rho^{-2/3}$ and $R \sim \rho$.
Thus the sensitivity scaling at high-densities is given by
\begin{align}
    \eta \propto \frac{\Gamma}{C_\textrm{r} \sqrt{R}} \propto \frac{\rho^{2/3}}{\rho^{-2/3} \sqrt{\rho}} = \rho^{5/6}
\end{align}
as stated in the main text (Fig.~3).
The scaling for typical electrometry protocols may be similarly determined.
In this case, all photons are scattered off-resonantly and  $C_\mathrm{r}$ is no longer relevant in the sensitivity expressions.
Then
\begin{align}
    \eta \sim \frac{\rho^{2/3}}{\sqrt{\rho}} = \rho^{1/6}
\end{align}
as shown in Fig.~3 of the main text.

\subsection{Isotopically Purified Samples}
We note that for isotopically purified samples the relative contributions of $\eta_\mathrm{F}$ and $\eta_\Pi$ may be comparable (see Tab.~\ref{tab:sensitivity_param}). 
To illustrate this, let us consider the sensitivity of NV electrometry protocols assuming a $^{13}\mathrm{C}$ concentration of $50$ ppm ([$^{12}\mathrm{C}$] = 99.995\% purified) as utilized in recently optimized samples \cite{edmonds_generation_2020, barry_sensitivity_2020}; indeed, such isotopically purified samples have been of tremendous recent interest for applications to precision metrology \cite{edmonds_generation_2020}.
Assuming an excited state broadening of $10$ GHz, we find that $\eta_\mathrm{F} = 1.4\cdot 10^{-3} \si{\volt/\centi\meter \sqrt{\hertz}}$ and $\eta_\Pi = 1.8\cdot 10^{-3} \si{\volt/\centi\meter \sqrt{\hertz}}$ at the optimal NV density $\rho_\mathrm{NV} \approx 2$ ppb.

\subsection{$^3 \rm{E}$ Fine Structure}
Our sensitivity model assumes that electric-field induced splitting dominates the $^3\rm{E}$ fine structure.
Under this assumption, it is sufficient to consider two three-fold degenerate orbital degrees of freedom in determining the resonance conditions.
However, this assumption breaks down when the splitting due to internal electric fields of the diamond becomes comparable to the intrinsic hyperfine effects of the $^3\rm{E}$ manifold, whose typical magnitude is $\sim 5 \si{\giga \hertz}$ \cite{maze_properties_2011}. 
Based on our extracted susceptibilities, we estimate this occurs for NV densities $\lesssim 2 \rm{ppb}$. 
For densities below this threshold, it is still possible to perform resonant sensing using individual sublevels; in particular, one should probe a pair of sublevels ($E_{x,y}$ or $E_{1,2}$) which are linearly sensitive to electric fields \cite{maze_properties_2011}.
In this regime, the sensitivity would be determined entirely by the change in overall fluorescence, i.e.~$\eta_\textrm{F}$.

\section{Theoretical Estimates of Susceptibilities} \label{app:suscep-theory}
In this section, we discuss the physical origin of the NV's electric field susceptibility in both the ground and excited state, and compare the measured susceptibility parameters to theoretical estimates (Table \ref{tab:susc_theory}).

\subsection{Excited state}
While it is well understood that the orbital doublet nature of the excited state allows for a linear Stark shift, the microscopic origin of this shift can in principle be explained by two different mechanisms \cite{maze_properties_2011}.
One mechanism, the electronic effect, is based on the polarization of the NV's electronic wavefunction.
The second mechanism, the ionic effect, consists of the relative displacement of the ions and is thus closely related to piezoelectricity.
The two effects are indistinguishable from a group theoretic perspective and, in general, both will contribute to the total susceptibility.
Below we estimate the susceptibilities based on the electronic effect and find good agreement with the measured values (Table  \ref{tab:susc_theory}).
%
%
%
%
On the other hand, the ionic effect was previously estimated with \emph{ab initio} simulations and was found to be on the same order of magnitude \cite{maze_properties_2011}.
Thus, a more precise calculation of the excited susceptibilities should take into account both effects.

\begin{table}
\centering
\begin{tabular}{c|c|c}
 \hline \hline
   \multirow{2}{*}{} & $\chi_\perp$  & $\chi_\parallel$ \\
   \multirow{2}{*}{} & (Hz cm/V) & (Hz cm/V) \\
   \hline
  E.S. measured (this work) & \makecell{$1.4\pm0.1 \times10^6 $} & \makecell{$0.7\pm0.1 \times10^6$}\\
  E.S. electronic effect & \makecell{$1.6 \times10^6$} & \makecell{$0.6\; \times10^6$}\\
  \hline
  G.S. measured \cite{van1990electric} & \makecell{$17\pm2.5$} & \makecell{$0.35\pm0.02$}\\
  G.S. spin-spin effect & \makecell{$76$} & \makecell{$0$}\\
  \hline \hline
\end{tabular}
\caption{Comparison between measured susceptibilities and theoretical estimates. Details of the theoretical estimates are provided in this work and accompanying references.}
\label{tab:susc_theory}
\end{table}

To estimate the electronic effect, we consider the molecular model of the defect center, in which the NV's single-particle orbitals are constructed from non-overlapping atomic orbitals, $\{\sigma_1,\sigma_2,\sigma_3 ,\sigma_N\}$, centered on the three carbon ions and the nitrogen ion, respectively \cite{doherty_negatively_2011,maze_properties_2011}.
In particular, the single-particle orbitals are given by
\begin{align}
e_x &= \frac{1}{\sqrt{6}} \left(2\sigma_1 - \sigma_2 - \sigma_3\right) \\
e_y &= \frac{1}{\sqrt{2}}\left(\sigma_2 - \sigma_3\right)\\
a_1 &= \frac{1}{\sqrt{3+\lambda^2}} \left(\sigma_1 + \sigma_2 + \sigma_3 + \lambda \sigma_N\right)\ ,
\end{align}
where $\sigma_1$ is the carbon orbital that lies in the $xy$ plane, and $\lambda \approx 0.7$ is determined from density functional theory (DFT) calculations \cite{gali2008ab}\footnote{We note that there is a fourth state with the same symmetry properties as $a_1$ (i.e.~transforms according to the totally symmetric $A_1$ irreducible representation), but it is higher in energy than the other three and therefore not relevant for this discussion~\cite{maze_properties_2011}.}
These orbitals are combined to form the $^3A_2$ ground state
\begin{equation}
    \ket {A_2} = \frac{1}{\sqrt 2}\left(\ket{e_x e_y} - \ket{e_y e_x}\right)
\end{equation}
and the two $^3E$ excited states
\begin{align}
    \ket {X} &= \frac{1}{\sqrt 2}\left(\ket{a_1 e_x} - \ket{e_x a_1}\right) \\
    \ket {Y} &= \frac{1}{\sqrt 2}\left(\ket{a_1 e_y} - \ket{e_y a_1}\right).
\end{align}

From first-order perturbation theory, the electric field susceptibility is determined by the permanent dipole of the excited state.
For the transverse susceptibility, it is sufficient to calculate the dipole moment along the $x$-axis, which is diagonal in the $\{\ket X, \ket Y\}$ basis:
\begin{align}
d_\perp &= -e\bra X x_1 + x_2 \ket X \ ,
\end{align}
where $ x_{1,2}$ are the single particle positions and $e$ is the elementary charge.
In the single-particle basis, this reduces to
\begin{align}
\abs{d_\perp} &= e\bra {e_x}  x \ket {e_x} \\
&\approx \frac{e}{2} \bra {\sigma_1} x \ket{\sigma_1}\ ,
\end{align}
where we have approximated the full integral by assuming non-overlapping atomic orbitals.
For the longitudinal direction, the relevant term is the relative dipole moment between between the ground and excited state. This is given by
\begin{align}
d_\parallel &= -e \left[\bra X  z_1 +  z_2 \ket X - \bra {A_2}  z_1 +  z_2 \ket {A_2}\right] \\
&= -e \left[\bra {a_1} z \ket {a_1} - \bra {e_x}  z \ket {e_x}\right] \\
&\approx \frac{e\lambda^2}{3+\lambda^2} \left[\bra {\sigma_1} z \ket{\sigma_1}- \bra {\sigma_N} z \ket{\sigma_N}\right].
\end{align}
Inserting orbital expectation values from DFT calculations, we obtain $\abs{d_\perp} \approx e(0.67~\si{\angstrom})$ and $d_\parallel \approx e(0.26~\si{\angstrom})$ \cite{gali2008ab,doherty_electronic_2014}.
This yields susceptibility estimates of $\{ \chi^\textrm{e}_\perp,\chi^\textrm{e}_\parallel\}  = \{1.6,0.6\}$ MHz/(V/cm), in good agreement with the values measured in this work (Table \ref{tab:susc_theory}).

\subsection{Ground state}
The ground state of the NV center is an orbital singlet, leading to the naive expectation that a linear Stark effect is disallowed.
This, however, contradicts experimental observation of $\{\chi^\textrm{g}_\perp,\chi^\textrm{g}_\parallel\}  = \{17,0.35\}~\si{\hertz / (\volt / \centi \meter)}$ \cite{van1990electric}.
The conventional explanation is that the ground state inherits a permanent dipole moment from the excited state due to spin-orbit coupling \cite{dolde_electric-field_2011,doherty2014measuring}.
While such coupling is indeed present, its magnitude is likely insufficient to account for the measured transverse field susceptibility.
More recently, it was suggested that the ground state transverse susceptibility arises from the interplay between electric fields and the dipolar spin-spin interaction \cite{doherty2014measuring}.
In particular, the effect is as follows:
At first order in perturbation theory, the ground state wavefunction is mixed with the excited state by the presence of an electric field; this perturbation then couples to the ground-state spin degrees of freedom via the dipolar spin-spin interaction.
Below we estimate the magnitude of the effect (which was not reported in \cite{doherty2014measuring}) and find good agreement with the known ground state transverse susceptibility (Table \ref{tab:susc_theory}).
We also hasten to emphasize that this effect only occurs to leading order for transverse electric fields, which naturally explains the 50-fold anisotropy between $\chi^\textrm{g}_\perp$ and $\chi^\textrm{g}_\parallel$.

As in the case of the excited state, it is sufficient to consider the transverse susceptibility for a field along the $x$-axis.
At first order in perturbation theory, an electric field $\vec E = E_\perp \hat x$ mixes the ground state $\ket {A_2}$ with the excited state $\ket {Y}$:
\begin{equation} \label{eq:pert_A2}
    \ket{A^\prime_2} = \ket{A_2} + \frac{E_\perp }{\nu_0}d_\perp^\prime \ket{Y},
\end{equation}
where $e$ is the elementary charge, and $\nu_0 \approx 1.9~\si{\electronvolt}$  is the energy splitting between the ground and excited state.
$d_\perp^\prime$ is the dipole moment associated with the transition between the states,
\begin{equation}
    d_\perp^\prime = -e\bra{A_2} x_1 +  x_2\ket{X}.
\end{equation}
In the single-particle basis, this becomes
\begin{align}
    \abs{d_\perp^\prime} &= e\bra{e_x} x \ket{a_1} \\
    &\approx \frac{3e}{\sqrt{6(3+\lambda^2)}} \bra{\sigma_1} x \ket{\sigma_1}.
\end{align}
Based on DFT results, we estimate $\abs{d_\perp^\prime} \approx e(0.88~\si{\angstrom})$ \cite{gali2008ab,doherty2014measuring}.

To determine the effect on the ground-state spin degrees of freedom, it is then necessary to consider the dipolar spin-spin interaction, given by
\begin{align}
    H_{ss} &= \eta \frac{3 (\bm{S}\cdot \hat r_{12})(\bm{S}\cdot \hat r_{12})-\bm{S}\cdot \bm{S}}{r_{12}^3}\ ,
\end{align}
where $\eta=\frac{ \mu_0 \mu_B^2 g^2}{8\pi h}$, $\mu_B$ is the Bohr magneton, $g_e \approx 2$ is the NV gyromagnetic ratio, $\bm{S}$ are spin-1 operators, and $\vec r_{12}$ is the relative displacement between the two particles.
In the absence of an external perturbation, the orbital degrees of freedom are integrated with respect to the ground-state wavefunction $\ket {A_2}$, and the only non-vanishing term is the ground-state splitting, $H^0_{ss} = \Delta_\textrm{ZFS} S_z^2$.
For the perturbed wavefunction $\ket{A_2^\prime}$, there is an additional non-vanishing term, corresponding to a ground-state Stark shift:
\begin{equation}
    H^\prime_{ss} = \Delta_\textrm{ZFS} S_z^2 + \Pi_\perp (S_y^2-S_x^2).
\end{equation}
The magnitude of $\Pi_\perp$ is given by
\begin{align}
    \Pi_\perp = 2\frac{E_\perp}{\nu_0}\abs{d_\perp^\prime} D_E
\end{align}
with
\begin{align}
D_E &= \eta \bra{A_2} \frac{x_{12}^2 - y_{12}^2}{r_{12}^5} \ket{Y} \\
&= \eta \bra{a_1 e_y} \frac{x_{12}^2 - y_{12}^2}{r_{12}^5} \left(\ket{e_x e_y} - \ket{e_y e_x}\right).
\end{align}
\begin{widetext}
Assuming non-overlapping orbitals, this simplifies to
\begin{align}
    D_E &\approx \frac{2 \eta}{\sqrt{6(3+\lambda^2)}} \left(\bra{\sigma_1 \sigma_2} \frac{x_{12}^2 - y_{12}^2}{r_{12}^5} \ket{\sigma_1 \sigma_2} + \bra{\sigma_2 \sigma_3} \frac{x_{12}^2 - y_{12}^2}{r_{12}^5} \ket{\sigma_2 \sigma_3} \right).
\end{align}
We further approximate the two-particle integrals with the semiclassical position of each particle individually \cite{doherty2014measuring,doherty_electronic_2014}:

\begin{align}
    \bra{\sigma_1 \sigma_2} \frac{x_{12}^2 - y_{12}^2}{r_{12}^5} \ket{\sigma_1 \sigma_2} &\approx \frac{\left(\left<x\right>_1-\left<x\right>_2\right)^2-\left(\left<y\right>_1-\left<y\right>_2\right)^2}{\left(\left<r\right>_1-\left<r\right>_2\right)^5} =\frac {1}{2\left(\left<r\right>_1-\left<r\right>_2\right)^3} = \frac{1}{6\sqrt{3}\left<x\right>_1^3} \\
    \bra{\sigma_2 \sigma_3} \frac{x_{12}^2 - y_{12}^2}{r_{12}^5} \ket{\sigma_2 \sigma_3} &\approx \frac{\left(\left<x\right>_2-\left<x\right>_3\right)^2-\left(\left<y\right>_2-\left<y\right>_3\right)^2}{\left(\left<r\right>_2-\left<r\right>_3\right)^5} =\frac {1}{\left(\left<r\right>_1-\left<r\right>_2\right)^3} = \frac{1}{3\sqrt{3}\left<x\right>_1^3}\ ,
\end{align}
\end{widetext}
where $\left<\cdot\right>_i = \bra{\sigma_i}\cdot\ket{\sigma_i}$ and in the final expressions we utilized the triangular symmetry of the carbon orbitals.
This leads to
\begin{equation}
    D_E \approx \frac{ \mu_0 \mu_B^2 g_e^2}{8 \pi h\sqrt{2(3+\lambda^2)}} \frac{1}{\left<x\right>_1^3}.
\end{equation}
Altogether, this predicts a susceptibility of $\chi^\textrm{e}_\perp  \approx 76~\si{\hertz / (\volt / \centi \meter)}$, which is within a factor of 5 of the measured value.

Crucially, the spin-spin effect also provides a group theoretic explanation for the large anisotropy between the ground state transverse and longitudinal susceptibilities.
In particular, the longitudinal dipole moment between the ground and excited states,
\begin{equation}
    d_\parallel^\prime = -e\bra{A_2} z_1 +  z_2\ket{X},
\end{equation}
vanishes due to symmetry, implying that \emph{only} a transverse electric field can mix the ground and excited states to leading order.
We thus postulate that the relatively strong transverse susceptibility arises from the proposed spin-spin effect, while the weak longitudinal effect arises from entirely different physical origin, e.g.~based on spin-orbit coupling or the ionic (piezoelectric) effect.

\bibliographystyle{unsrt}
\bibliography{bibliography.bib}

\begin{thebibliography}{100}
\bibitem{7_foonote_config}
We define a resonant configuration, at a given detuning, as a combination of
  transverse and longitudinal electric fields which cause the NV's ground to
  excited state transition to be resonantly driven by the optical excitation.
  Note that the resonance condition does not depend on the direction of the
  transverse component of the field in the plane perpendicular to the NV axis.
\bibitem{!_footnote_widthlim}
Other perturbations to the NV level structure, such as strain and magnetic
  fields, can exacerbate ensemble broadening. For this discussion, we restrict
  attention to the ideal limit where there is minimal charge ($\rho_c =
  2\rho_\mathrm{NV}$) and there are no other significant density dependent
  sources of inhomogenous broadening.
\end{thebibliography}

\end{document}